\documentclass[fleqn,usenatbib]{mnras}

\usepackage{newtxtext,newtxmath}

\usepackage[T1]{fontenc}

\DeclareRobustCommand{\VAN}[3]{#2}
\let\VANthebibliography\thebibliography
\def\thebibliography{\DeclareRobustCommand{\VAN}[3]{##3}\VANthebibliography}

\usepackage{gensymb}    
\usepackage{graphicx}	
\usepackage{amsmath}	
\usepackage{comment}    
\usepackage{rotating}   
\usepackage{hyperref}   
\usepackage{caption}
\usepackage{subcaption}
\usepackage{soul}
\usepackage[normalem]{ulem}
\usepackage{dirtytalk}
\usepackage[dvipsnames]{xcolor}
\usepackage{float}
\usepackage[dvipsnames]{xcolor}
\usepackage{graphicx}
\usepackage{orcidlink}
\usepackage{verbatim}
\newcommand{\ppxf}{{\sc pPXF }}

\title[Properties of Dual Nuclei Systems in GOTHIC]{Investigating the Spectral Properties of Dual Nuclei in Galaxy Mergers from the GOTHIC survey: Supermassive Black Hole Growth, metal enrichment and Dual AGN}

\author[Biswas et al.]{
Prerana Biswas$^{1}$\orcidlink{0009-0006-7375-6580} \thanks{E-mail: prerana.biswas@iiap.res.in},
Mousumi Das$^{1}$\orcidlink{0000-0001-8996-6474} \thanks{E-mail: mousumi@iiap.res.in},
Sudhanshu Barway$^{1}$\orcidlink{0000-0002-3927-5402},
Françoise Combes$^{2}$\orcidlink{0000-0003-2658-7893}, 
Anwesh Bhattacharya$^{3}$,
\newauthor
Snehanshu Saha$^{3,4}$\orcidlink{0000-0002-8458-604X},
C. P. Nehal$^{5}$\orcidlink{0009-0008-9053-7034}
\\
$^{1}$Indian Institute of Astrophysics, 2nd Block, Koramangala, Bengaluru - 560034, India\\
$^{2}$Observatoire de Paris, LERMA, College ' de France, PSL University, Sorbonne University, CNRS, Paris, France-75014\\
$^{3}$Siebel School of Computing and Data Science, University of Illinois at Urbana-Champaign, Urbana, IL 61801, USA\\
$^{4}$APPCAIR, Department of CSIS, Birla Institute of Technology \& Science and HappyMonk AI, Goa 403726, India \\
$^{5}$Department of Physics, Indian Institute of Science Education and Research, Bhopal 462066, India
}

\date{Accepted XXX. Received YYY; in original form ZZZ}

\pubyear{2015}

\begin{document}
\label{firstpage}
\pagerange{\pageref{firstpage}--\pageref{lastpage}}
\maketitle

\begin{abstract}
Dual nuclei systems are galaxy merger remnants or closely merging galaxies that have two distinct stellar cores separated by $\sim$10pc to 10kpc. They are important laboratories for probing the co-evolution of stellar populations, galaxy dynamics, and central black holes during the hierarchical assembly of galaxies. In this study, we present a spectroscopic analysis of a sample of dual nuclei from the GOTHIC survey, using the penalized pixel-fitting (pPXF) code. The sample consists of star forming nuclei pairs, dual active galactic nuclei (DAGN) and mixed pairs. Using the SDSS spectra, we extracted stellar kinematics, emission line fluxes, the star formation history, metallicity of the nuclei, and derived important properties such as the supermassive black hole (SMBH) masses, accretion rates and SMBH ratios. We compared different properties of the nuclei in the dual systems, such as stellar velocity dispersion, stellar masses, black hole masses, age and metallicity. Our results show that the SMBH masses are higher for BHs in galaxy mergers compared to single nuclei for a given stellar mass, thus revealing that SMBHs grow during the galaxy merging process and not only due to the merger of SMBHs. Our study provides new observational constraints on the dynamical and evolutionary states of dual-nuclei systems, offering a deeper understanding of the role these systems play in galaxy evolution and central black hole growth.

\end{abstract}



\section{Introduction}
\label{sec:intro}

Galaxy mergers play a fundamental role in shaping galaxy evolution \citep{hierarchicalgrowth_White_Frenk1991, gal_merger_evo_Beckman2008}, driving processes such as gas inflows, starbursts, and supermassive black hole (SMBH) growth, often leading to the morphological transformation of galaxies \citep{martin.etal.2018}. During these interactions, it is common for galaxies to host more than one luminous nucleus, 
that are powered by strong central activity such as star formation, active galactic nuclei (AGN) or both \citep{yadav.etal.2023}. Such multi-nuclei or dual-nuclei systems represent dynamically complex stages of galaxy assembly where two (or more) central stellar concentrations coexist  either within a common envelope or at close separations \citep{Das.etal.2018}. Hence they can be roughly classified to be merging galaxies where the nuclear separations lie between a few tens of kpc to $\sim$100 pc. Binary nuclei, however, are separated by $<$ 100pc.  Dual and binary nuclei are important for understanding the intermediate phases of supermassive black hole (SMBH) pairing before the SMBH inspiral and final coalescence \citep{enoki.etal.2004,khan.etal.2018}.

Studying multi-nucleus systems is important for several reasons. First, they allow us to trace the co-evolution of galaxies and their central black holes during the merger process, a regime in which accretion, star formation, and chemical enrichment may be strongly enhanced or disrupted depending on the merger parameters \citep[e.g., see ][]{Heckman2014_coevl_smbh4, DEROSA2019_coevl_smbh3, Mountrichas2023_coevl_smbh1, Li2023_coevl_smbh2}. Second, they provide constraints on scaling relations, such as the black hole mass–stellar mass relation \citep{Ferrarese2000, Kormendy_Ho_2013} or the stellar mass–velocity dispersion relation \citep{sig_mstar_Zahid2016, mzr_Gallazzi2005, portsmouth_thomas_2013}, under conditions where dynamical equilibrium may not be fully established \citep[e.g., see][]{Barnes1992_instab_merger1, Jesseit2007_instab_merger2, Ellision2013_instab_merger3}. Third, because the two nuclei share the same global gravitational potential but may experience very different local gas environments, they offer a direct way to test whether star-formation histories, AGN triggering, or chemical evolution proceed synchronously or evolve independently \citep{rubinur.etal.2019}. Despite their importance, multi-nucleus systems in large spectroscopic surveys such as SDSS have not been systematically isolated and analysed as a distinct population, leaving several open questions regarding their physical properties and evolutionary pathways \citep{deRosa.etal.2019}.

In this study, we aim to fill this gap by using multi-nuclei galaxies from SDSS and performing a detailed investigation of the spectral properties of the individual nuclei. Our goals are to (i) determine whether galaxies in multi-nuclei systems follow the same global scaling relations as isolated systems, (ii) explore whether paired nuclei co-evolve or display divergent stellar population histories, and (iii) quantify how black hole mass, metallicity, stellar age, and velocity dispersion compare between the two components of each system. We also examine how dual systems populate different merger categories, such as major or minor mergers, and investigate what factors SMBH growth and AGN activity depend on  \citep{treister.etal.2010,lin.etal.2023}.

In this paper we use a sample of dual and multiple nuclei derived from the SDSS DR18 catalogue by \citet{gothic_2023}, also called the GOTHIC sample, to address the above questions. Further inspection of this sample shows that dual AGN are more likely to form in red cloud, evolved galaxies. Follow-up examination of the bulges of the dual AGN host galaxies indicates that in $\sim$2/3 of the galaxy pairs at least one galaxy is an elliptical galaxy, and $\sim$1/3 of the dual AGN are elliptical-elliptical merging pairs \citep{nehal.etal.2025}. So dual AGN are more likely to be found in red, quenched galaxies, that are major mergers. In this study we take this a step further and examine the nuclei of the merging galaxy sample, including both star forming, dual AGN and mixed pairs.

The structure of this paper is as follows. Section~\ref{sec:data} describes the dataset, while Section~\ref{sec:analysis} outlines the spectral fitting methodology. The main results are presented in Section~\ref{sec:results} and discussed in Section~\ref{sec:discussion}. Finally, Section~\ref{sec:conclusion} summarises the key findings and presents our conclusions.

\section{Sample selection and data}
\label{sec:data}
For our study of multi-nuclei galaxies,we use the sample of multi-nuclei sources from \cite{gothic_2023}, where a novel algorithm called GOTHIC (Graph-bOosTed iterated HIll Climbing), is applied to SDSS images to find galaxies where two or more closely separated nuclei are observed. The GOTHIC algorithm was applied to a sample of one million randomly chosen SDSS  \citep{sdss_iv_overview} galaxies from DR16 {\footnote{\url{https://www.sdss4.org/dr16/}}.After the automated detection and several filtration steps, 949 candidates of multi-nuclei systems were identified. As GOTHIC sometimes detects false positive results, the 949 candidates were manually investigated, and 681 multi-nuclei systems were confirmed. This process ensured a robust, spectroscopically confirmed sample of dual and multi-nuclei galaxies from SDSS-DR16. 

The final sample of 681 multi-nuclei systems consists of 1393 galaxies, among which 652 are dual-nuclei systems, 27 are triple-nuclei systems, and 2 are quadruple-nuclei systems. The finalised sources comprise both brighter and relatively fainter sources, with their SDSS r-band magnitude mostly ranging from $\sim$15 to $\sim$20. The redshift (z) distribution of these sources is primarily between z = 0 and 0.3, with a peak around z $\sim$ 0.073 \citep[see figure 3b from][]{gothic_2023}. Furthermore, most of these galaxies are found in the redder and more massive region of the colour-magnitude diagram (CMD), indicating they are typically massive galaxies with low recent star formation \citep[see figure 6 from][]{gothic_2023}. This final sample consists of star-forming galaxies, active galactic nuclei (AGN) and composite galaxies. Further, it is to be noted that the spectroscopic data for all of these sources were already available from SDSS \citep{sdss_iv_overview} and we made use of these spectra for deriving the different properties of these multi-nuclei galaxies.

There are other studies in the literature that have provided large catalogues of SDSS galaxies and their spectra. Notably, the FIREFLY catalogue of \citet{Firefly_wlikinson_2017} performed full spectral fitting to derive stellar population parameters using grid-based $\chi^2$ minimisation over combinations of SSP models. FIREFLY does not enforce smoothness constraints on the recovered star formation histories. In contrast, the {\sc pPXF} method used in this study applies a penalised $\chi^2$ formalism, combining linear and non-linear least-squares fitting with optional regularisation that encourages smooth star formation histories. Furthermore, FIREFLY estimates properties such as stellar mass and velocity dispersion using PCA-based approaches, while {\sc pPXF} derives them from template-weighted contributions. We find that all finalised sources used in this work have matches within $3''$ of the FIREFLY catalogue, giving confidence in our source selection and enabling comparison with prior spectral fitting studies.

For completeness, we also cross-matched our sample with the other value added catalogues (\href{https://www.sdss.org/dr19/data_access/value-added-catalogs}{VAC}\footnote{\url{https://www.sdss.org/dr19/data_access/value-added-catalogs}}) of SDSS, where different pipelines fit spectra using a combination of stellar population synthesis templates and emission-line modelling to derive redshifts, velocity dispersions, emission-line fluxes, and other fundamental properties. The number of sources matching our finalised sample in these catalogues is very small. For instance, only 11 sources match within a $3^{\prime\prime}$ radius in the Portsmouth catalogue \citep{portsmouth_thomas_2013}.  From MaNGA-FIREFLY \citep{firefly_mass_diff_Neumann2022} catalogue we found that only 32 sources match our finalised sample within a $3^{\prime\prime}$ radius. Similarly, the Wisconsin catalogue \citep{wisconsin2012} contains 10 matching sources within a $3^{\prime\prime}$ radius, and the Granada group catalogue \citep{granada2009, granada2016} also yields only 10 matched sources. This indicates limited overlap between those catalogues and our final multi-nucleus sample and further motivates an independent analysis of the stellar populations and black holes in multi-nucleus systems using a homogeneous method such as {\sc pPXF}.


\begin{figure}
    \centering
    \begin{subfigure}[b]{0.49\textwidth}
        \centering
        \includegraphics[width=\textwidth]{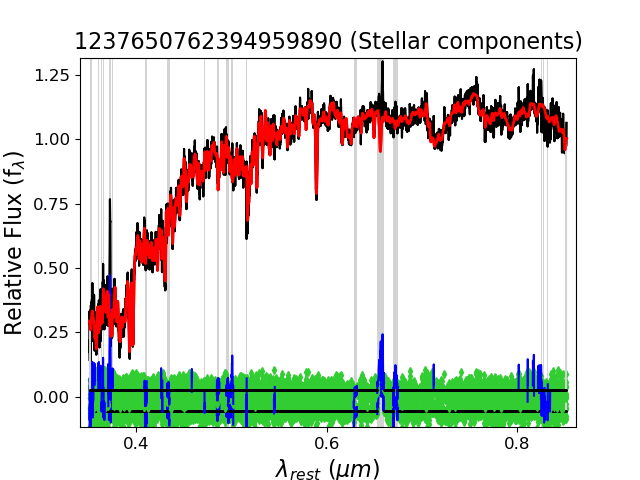}
        \caption{quality flag = 1}
        \label{fig:flag0}
    \end{subfigure}
    \hfill
    \begin{subfigure}[b]{0.49\textwidth}
        \centering
        \includegraphics[width=\textwidth]{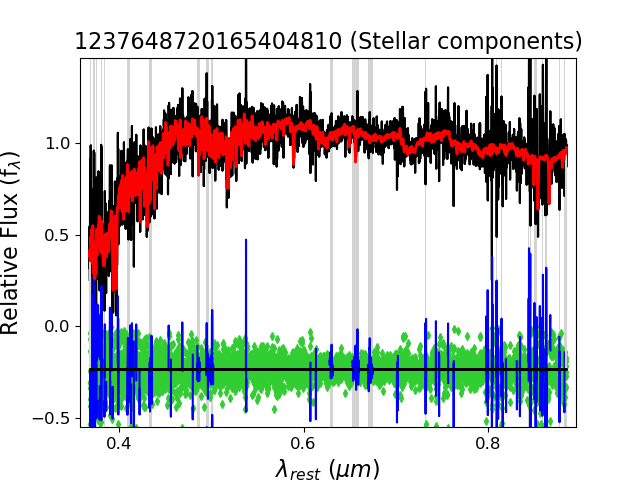}
        \caption{quality flag = 2}
        \label{fig:flag1}        
    \end{subfigure}
    \caption{Comparison of the observed spectra with fitted model using stellar components only from {\sc pPXF} for two sources categorised as quality flag is 1 \& 2, respectively. The black and red points, respectively, represent the observed spectra and the fitted spectra. Green points denote the residuals from the fit, whereas blue points indicate those residuals that exceed the 3$\sigma$ threshold of the residual distribution. The grey vertical lines mark the pixels identified and clipped as outliers.}
    \label{fig:ppxf_fit_stellar}
\end{figure}

\begin{figure*}
     \centering
     \begin{subfigure}[b]{0.49\textwidth}
         \centering
         \includegraphics[width=\textwidth]{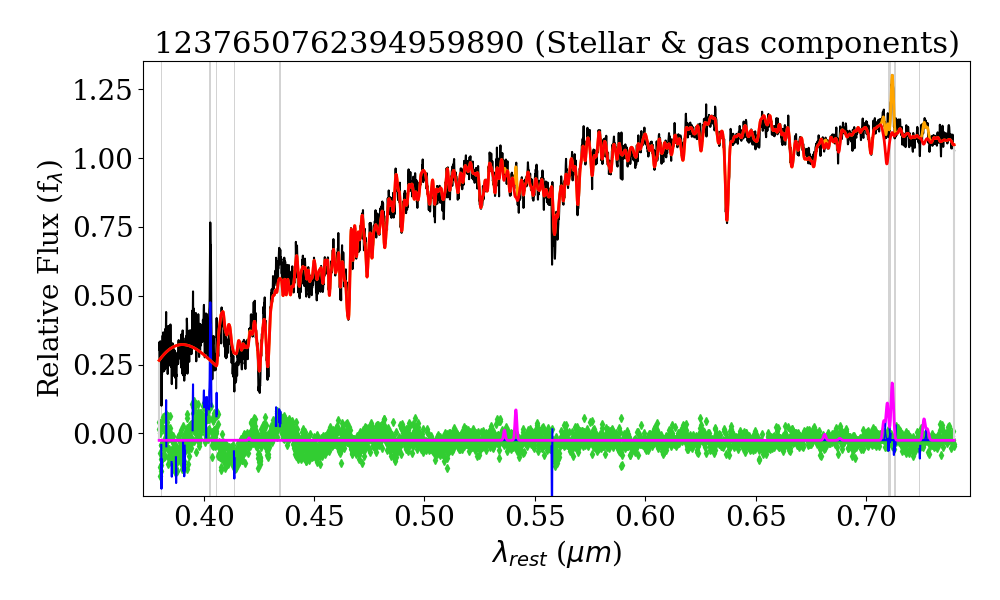}
     \end{subfigure}
     \hfill
     \begin{subfigure}[b]{0.45\textwidth}
         \centering
         \includegraphics[width=\textwidth]{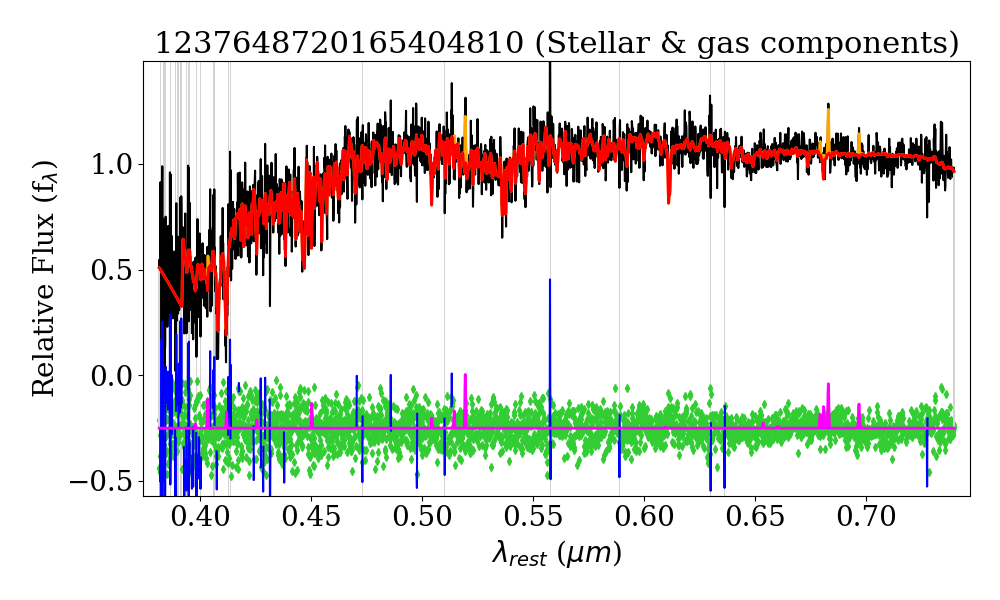}
    \end{subfigure}
    \hfill
    \begin{subfigure}[b]{0.45\textwidth}
         \centering
         \includegraphics[width=\textwidth]{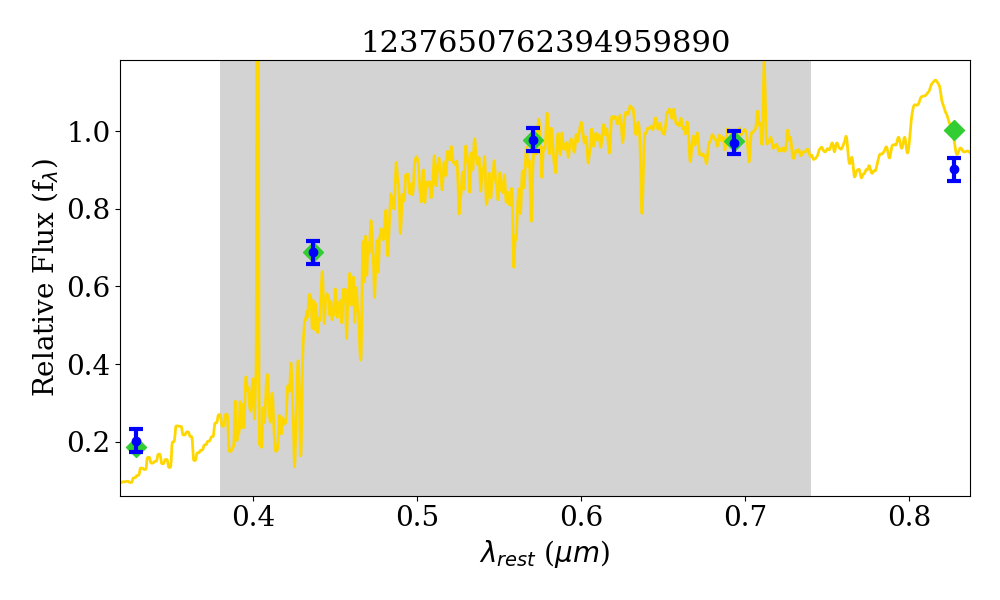}
         \caption{quality flag = 0}
         \label{fig:flag0_stellar_gas}
     \end{subfigure}
     \hfill
     \begin{subfigure}[b]{0.45\textwidth}
         \centering
         \includegraphics[width=\textwidth]{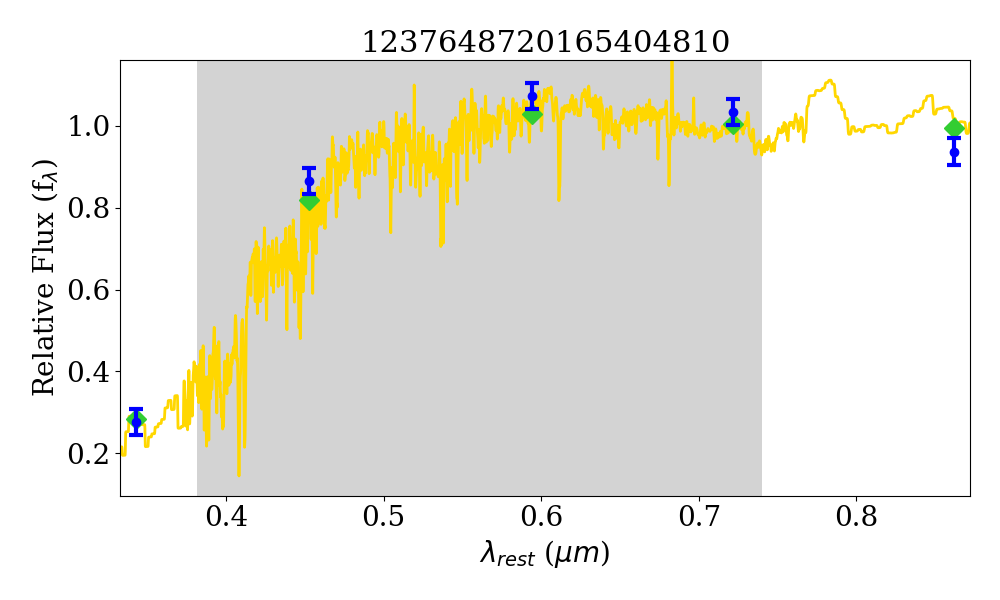}
         \caption{quality flag = 1}
         \label{fig:flag1_stellar_gas}
     \end{subfigure}
        \caption{Comparison of the observed spectra with fitted model considering both the gas and stellar emission obtained using {\sc pPXF} for two sources categorised as quality flag is 0 \& 1, respectively. Top panel: the black, orange and red lines respectively represent the observed spectra, the best-fitted spectra containing both gas and stellar contributions and the best-fitted stellar spectra. The magenta line denotes the best-fitted gas emission lines found in the {\sc pPXF}. Green points denote the residuals from the fit, whereas blue points indicate those residuals that exceed the 3$\sigma$ threshold of the residual distribution. The grey vertical lines mark the pixels identified and clipped as outliers. Bottom panel: Blue circles and green diamonds represent the observed SDSS fluxes and the best-fit fluxes in the $u, g, r, i,$ and $z$ bands, respectively. The yellow line shows the best‐fitting spectrum, and the grey shaded region indicates the wavelength range used in the fitting.}
        \label{fig:ppxf_fit_stellar_gas}
\end{figure*}

\begin{figure*}
    \centering
    \begin{subfigure}[b]{0.24\textwidth}
        \centering
        \includegraphics[width=\textwidth]{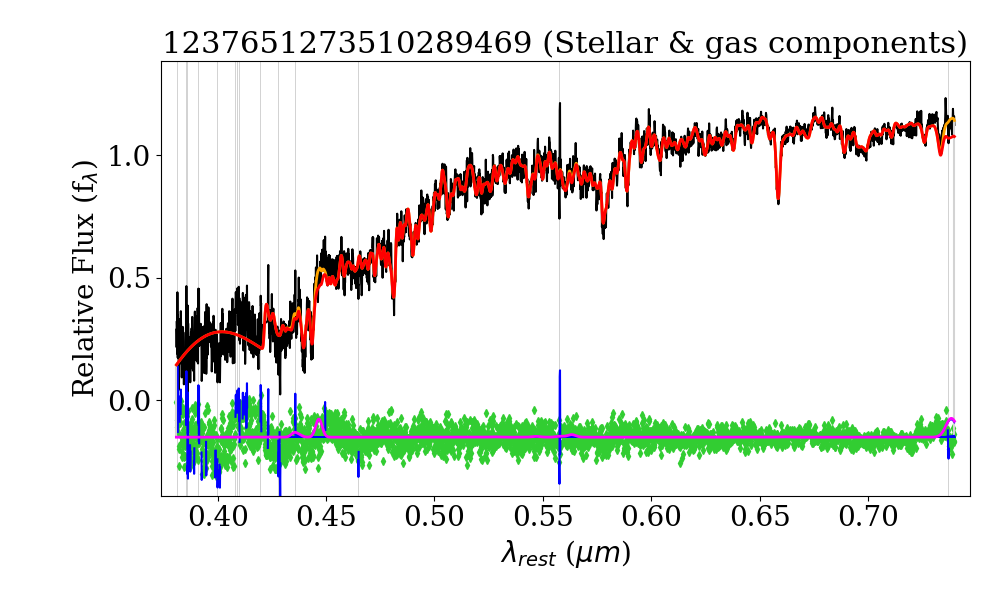}
    \end{subfigure}
    \hfil
    \begin{subfigure}[b]{0.24\textwidth}
        \centering
        \includegraphics[width=\textwidth]{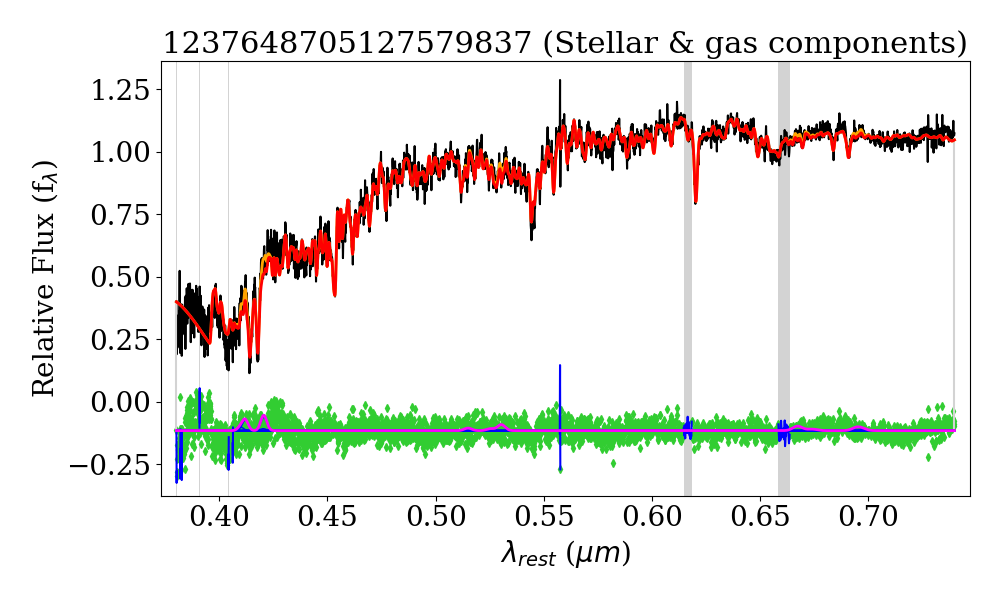}
    \end{subfigure}
    \hfil
    \begin{subfigure}[b]{0.24\textwidth}
        \centering
        \includegraphics[width=\textwidth]{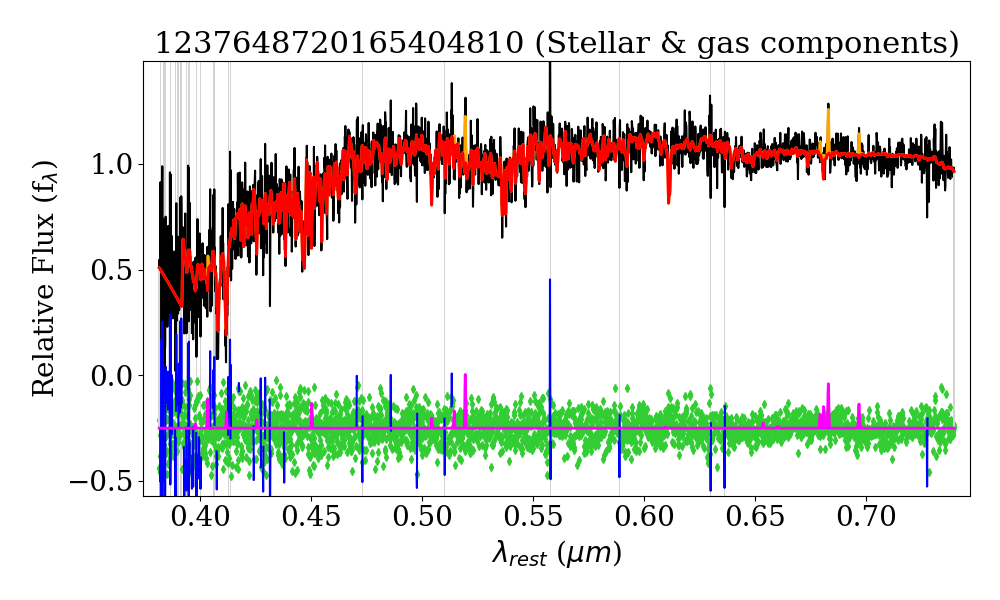}
    \end{subfigure}
    \hfil
    \begin{subfigure}[b]{0.24\textwidth}
        \centering
        \includegraphics[width=\textwidth]{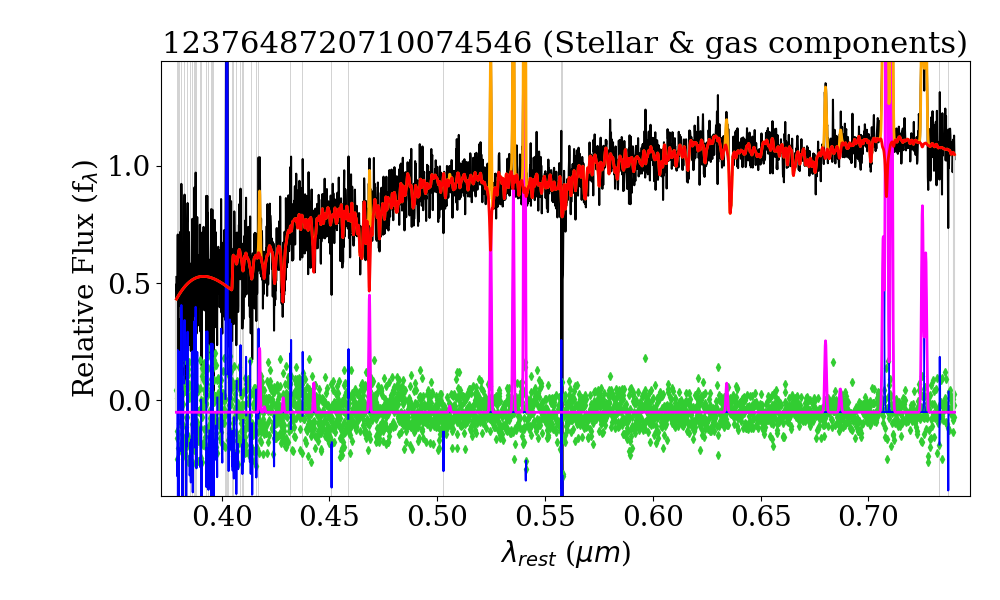}
    \end{subfigure}
    \hfil
    \begin{subfigure}[b]{0.24\textwidth}
        \centering
        \includegraphics[width=\textwidth]{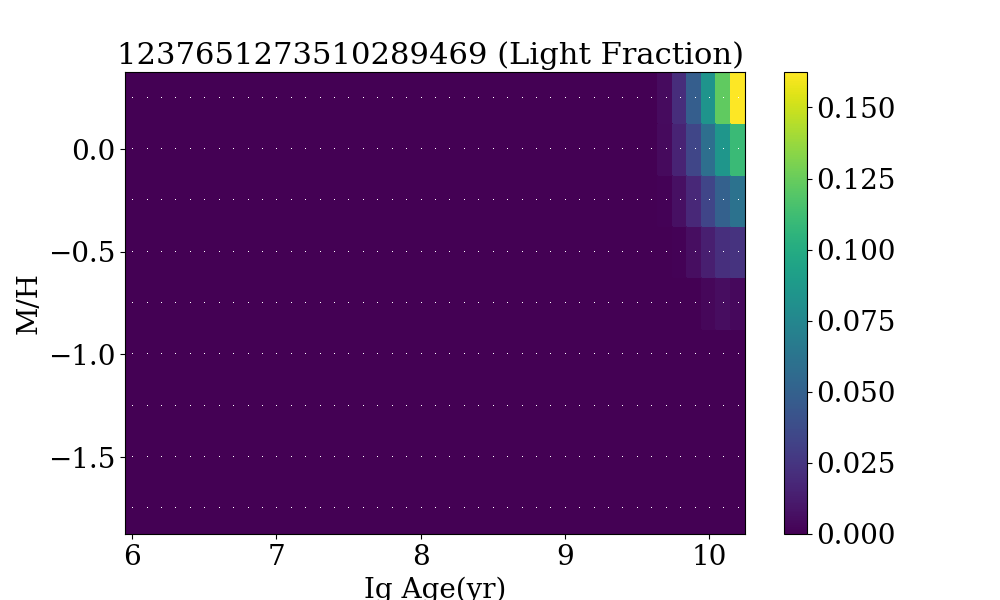}
        \caption{Star forming}
        \label{fig:sf_lightfrac}
    \end{subfigure}
    \hfil
    \begin{subfigure}[b]{0.24\textwidth}
        \centering
        \includegraphics[width=\textwidth]{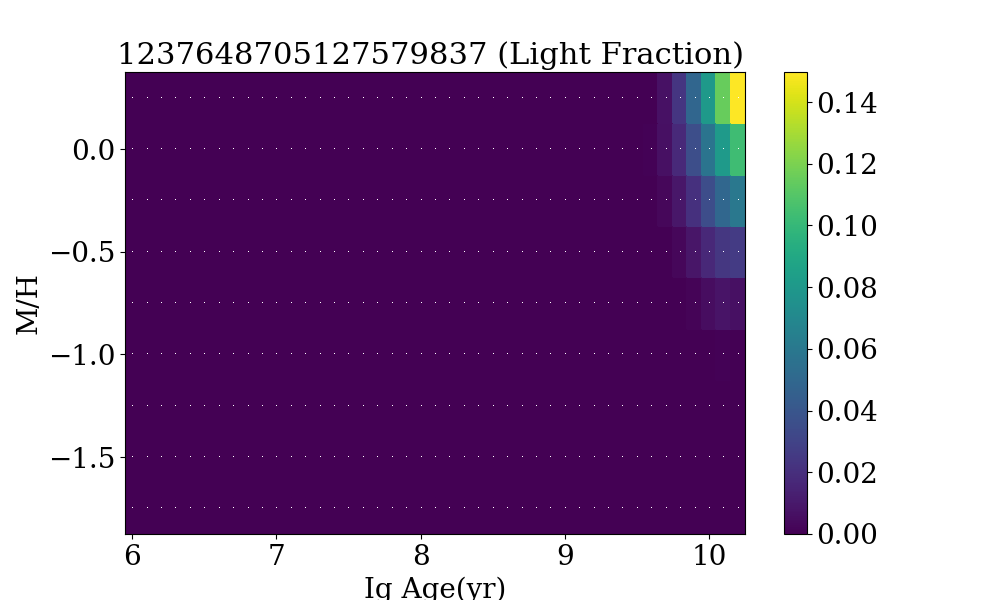}
        \caption{Composite}
        \label{fig:comp_lightfrac}
    \end{subfigure}
    \hfill
        \begin{subfigure}[b]{0.24\textwidth}
        \centering
        \includegraphics[width=\textwidth]{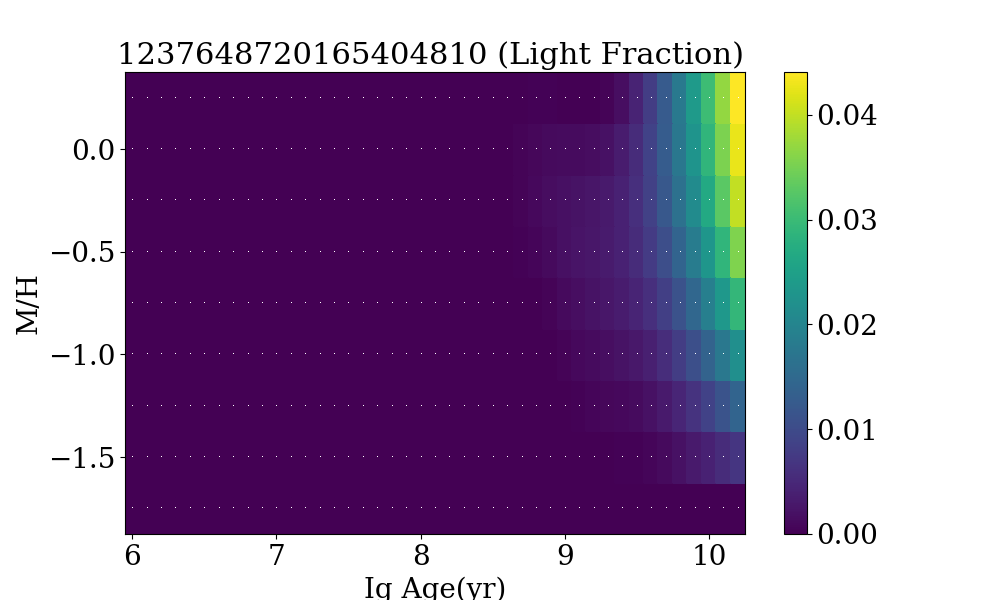}
        \caption{AGN-LINER}
        \label{fig:agnlinr_lightfrac}
    \end{subfigure}
    \hfill
        \begin{subfigure}[b]{0.24\textwidth}
        \centering
        \includegraphics[width=\textwidth]{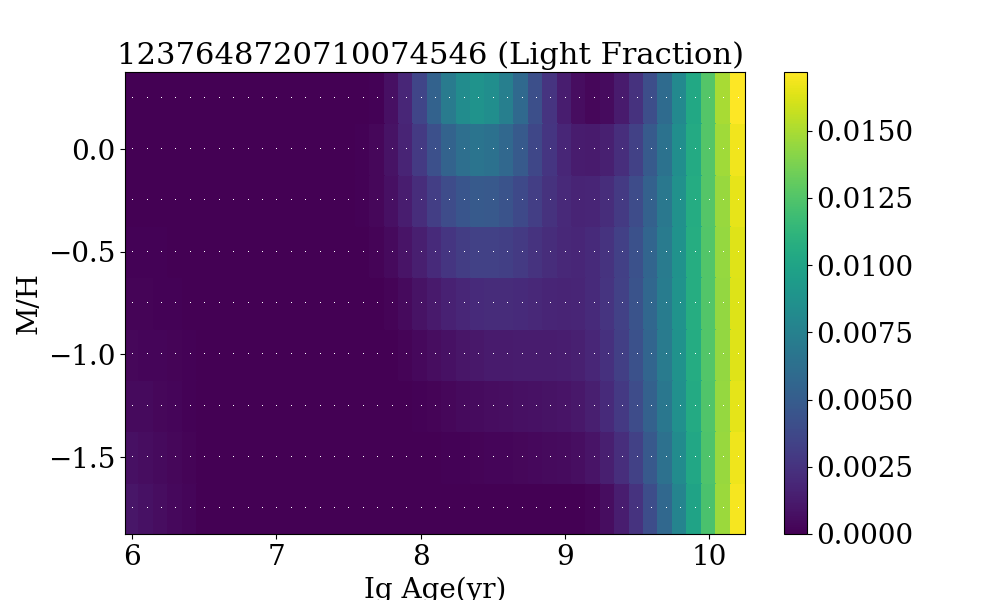}
        \caption{AGN-Seyfert}
        \label{fig:agn_seyfert_lightfrac}
    \end{subfigure}
\caption{Top panel: Observed and fitted spectra for representative galaxies classified as star-forming, Composite, AGN–LINER, and AGN–Seyfert in our sample. The black, orange and red lines respectively represent the observed spectra, the best-fitted spectra containing both gas and stellar contributions and the best-fitted stellar spectra. The magenta line denotes the best-fitted gas emission lines found in the {\sc pPXF}. Green points denote the residuals from the fit, whereas blue points indicate those residuals that exceed the 3$\sigma$ threshold of the residual distribution. The grey vertical lines mark the pixels identified and clipped as outliers. Bottom panel: Stellar metallicity versus stellar age for corresponding sources in the top panel. The colour scale represents the light fraction contribution from each stellar population component, as derived from the pPXF spectral fitting.}
\label{}
\end{figure*}

\section{Spectral fitting and analysis}
\label{sec:analysis}
To obtain reliable spectral fits and hence increase the accuracy of our results for the  GOTHIC sources, we checked the quality of the SDSS spectra by finding their signal-to-noise ratio (SNR). We use the Python package {\sc specutils} for finding the SNR of these sources. This package uses the robust DER\_SNR algorithm \citep{DER_SNR2008} to find the SNR of the spectra. We adopt a cutoff of SNR = 10 to filter out noisy spectra. Although some earlier studies of SDSS spectra \citep{SNR15} have used higher cutoffs (SNR$\sim$15), we choose to retain this more modest threshold. After filtering out the noisy spectra, among the 1393 finalised sources, we are left with 1040 sources. Our further analysis is based on these sources. 

For fitting the spectra and investigating the stellar and gas kinematics, and other properties, such as the line-flux of different lines present in the galaxy, the stellar mass, age, and metallically of these sources, we used the Penalized Pixel-Fitting stellar kinematics extraction ({\sc pPXF}) \citep{Cappellari_&_Emsellem_2004, Cappellari2017_ppxf, Cappellari2023_ppxf}. We fit the spectra using {\sc pPXF} in two different modes: first, by masking the gas emission line for a more accurate determination of the stellar kinematics, i.e., the stellar velocity dispersion and second, without masking the gas emission lines for determining other properties. For both cases, the same standard initial steps have been followed, i.e., the flux is normalised by its median value to improve numerical stability during fitting, the noise is determined by the mean of the noise provided in the spectral data and assumed to be constant across all pixels and stellar population synthesis (SPS) template, E-MILES library \citep{Emiles_2016} is used for covering a broad wavelength range,

\subsection{Stellar Kinematics}
\label{subsec:stellar_kin}
To obtain the stellar kinematics using {\sc pPXF}, the following procedures have been followed. We run the {\sc pPXF} twice: first with mean noise determined from the data and masking the gas emission lines by utilising the {\sc goodpixels} function of {\sc pPXF}. After the first run the noise is rescaled using the reduced $\chi^2$ found in the first run as follows: $ns_{re} = ns_{n}/\sqrt(\chi^2)$, where  $ns_{re}$ is the rescaled noise and $ns_{n}$ is the noise of the spectra determined from data as mentioned above. After that, we calculate the residuals of the fitting from the first run, and the outliers in the residuals are found using the robust determination of the sigma \cite{robust_sigma, Cappellari2017_ppxf}. The outlier points in the input spectra are then clipped. The {\sc pPXF} is then run for the second time with these clipped spectra, masked emission lines and rescaled noise for a robust fitting and determination of the stellar kinematics. 

\begin{figure*}
    \centering
    \begin{tabular}{cc}
         \includegraphics[width=0.4\textwidth]{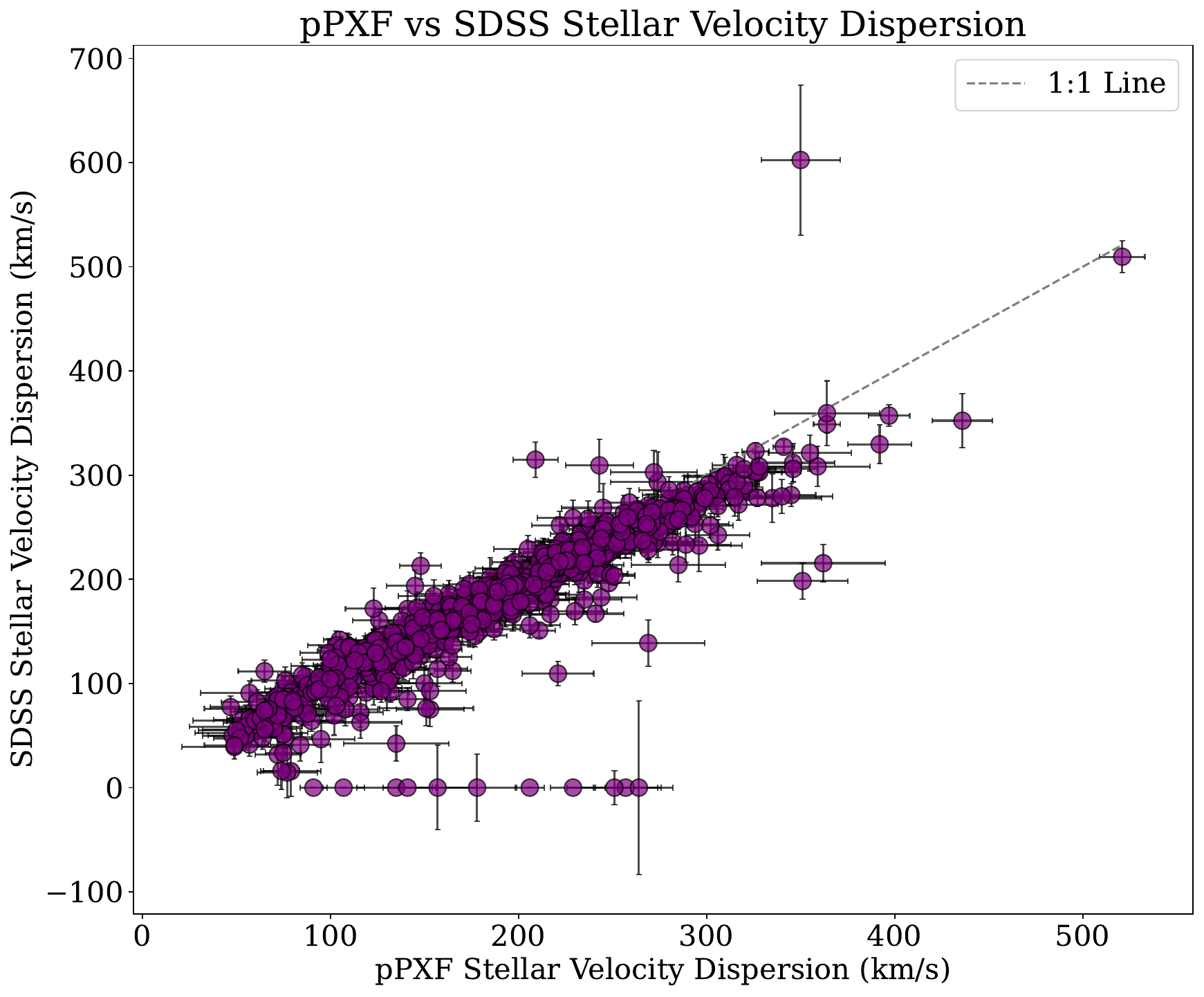} &
         \includegraphics[width=0.4\textwidth]{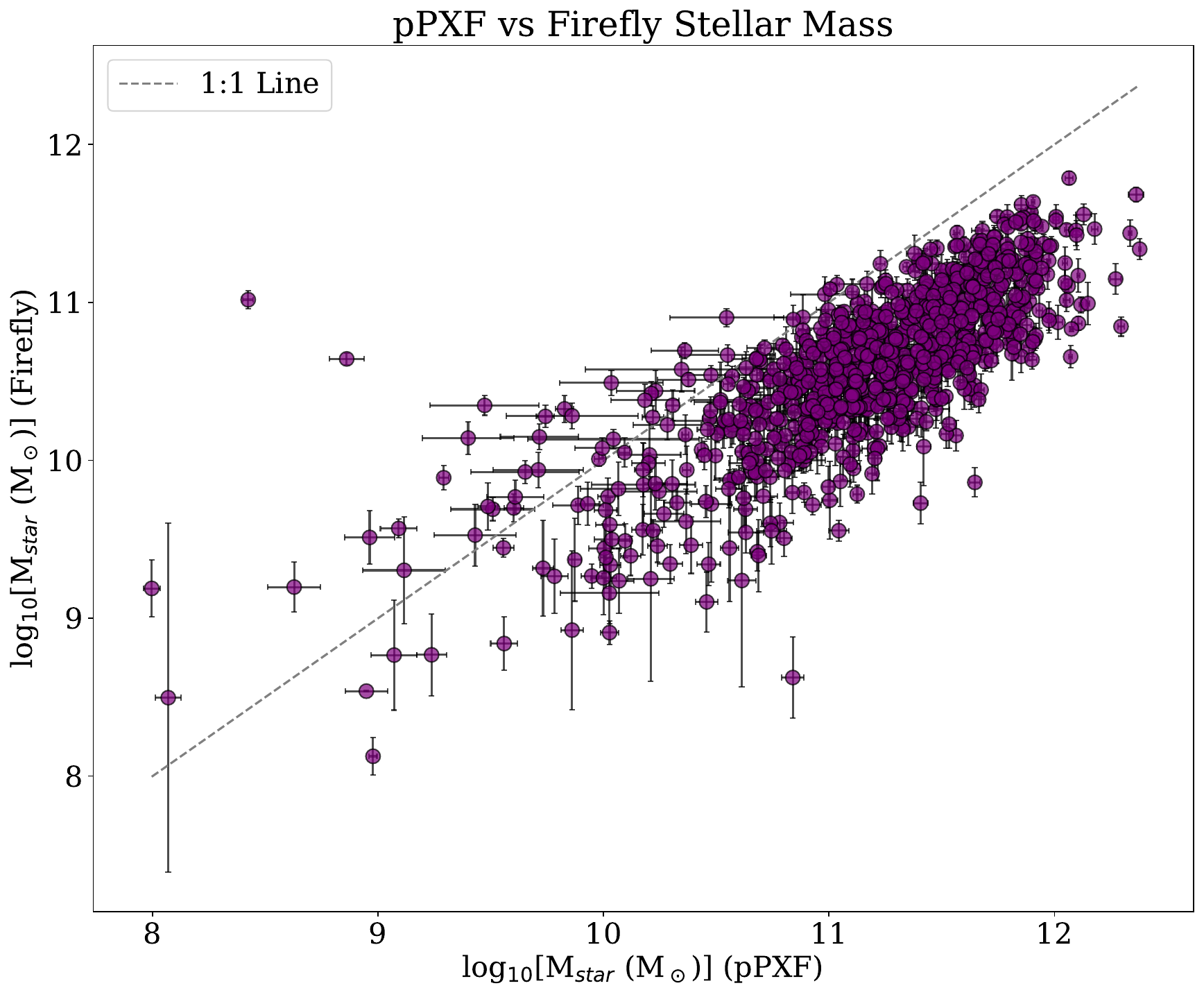} \\
         \includegraphics[width=0.4\textwidth]{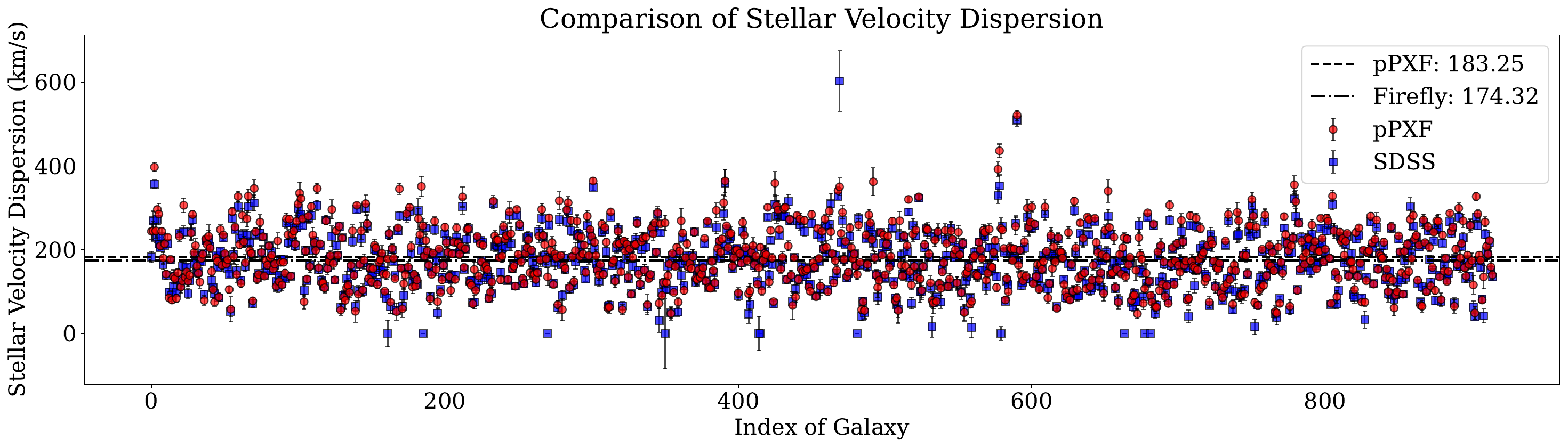} &
         \includegraphics[width=0.4\textwidth]{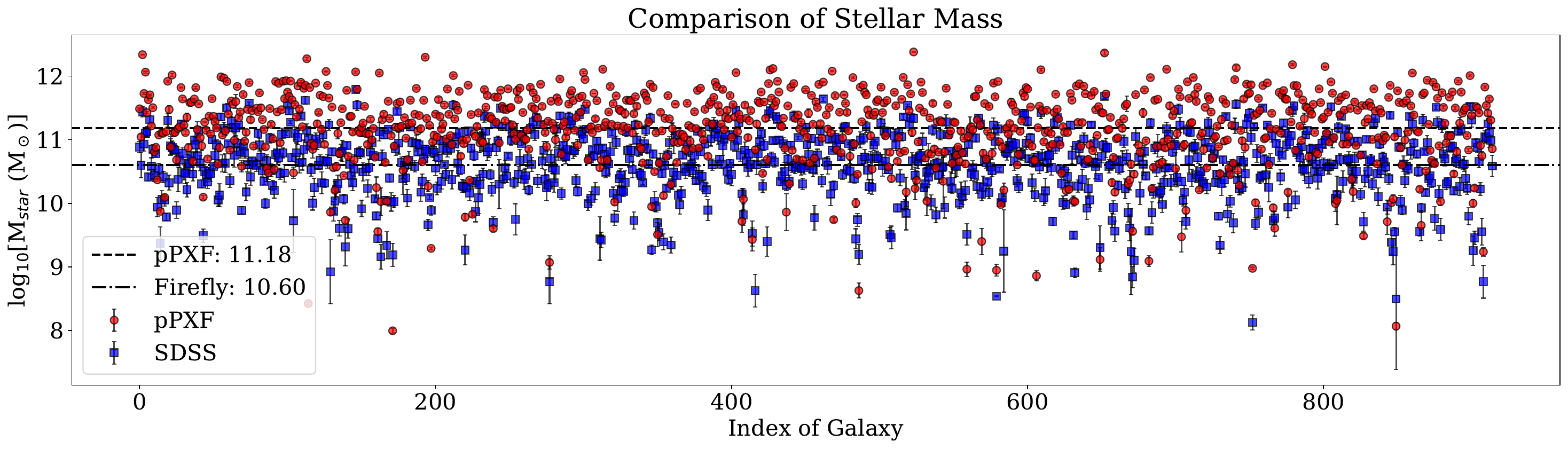} 
         
    \end{tabular}
    \caption{Left column: Comparison of stellar velocity dispersion from our analysis with measurements from SDSS. The black dashed line and dash-dotted lines in the bottom panel respectively show the average stellar dispersion velocity of all the finalised sources found in our analysis and from SDSS.  Right column: Comparison of the stellar masses from our analysis with the measurements from FIREFLY \citep{Firefly_wlikinson_2017}. The black dashed line and dash-dotted lines in the bottom panel respectively show the average stellar mass of all the finalised sources found in our analysis and from FIREFLY.}
    \label{fig:consitency_check}
\end{figure*}

\subsection{Identification of the acceptable fits}
\label{subsec:accptable_fits}

All 1,040 good-quality spectra were fitted using this process. However, the fit quality was not satisfactory for all cases. Therefore, we visually inspected all fittings and found that they form four visually distinct and unambiguous categories based on the quality of the fittings: good fit (qflag = 1), nearly good fit (qflag = 2), nearly bad fit (qflag = 3), and bad fit (qflag = 4). Examples of qflag = 1 and qflag = 2 spectra are shown in Figure \ref{fig:ppxf_fit_stellar}, while examples of qflag = 3 and qflag = 4 spectra are presented in Figure \ref{fig:ppxf_fit_stellar_bad} of Appendix \ref{sec:spec_bad}. A total of 915 sources, classified as good fits (qflag = 1) or nearly good fits (qflag = 2), were retained for further analysis.

\subsection{Inclusion of gas emission lines}
\label{subsec:gas_lines}
As noted earlier, after extracting the stellar kinematics by masking the gas emission lines, we again use {\sc pPXF} without masking these lines to determine the other properties of the sources. In this case also, we first run {\sc pPXF} to find the outliers from the fitting residuals using the robust sigma determination method as mentioned in subsection \ref{subsec:stellar_kin}. Then we mask these pixels in the input spectra corresponding to outliers in the residuals and re-run {\sc pPXF}. In this second run, we also normalised the noise as mentioned in subsection \ref{subsec:stellar_kin}. In both cases, i.e., fitting the spectra with masked gas emission lines and without masking them,  we regularise the weights during the fitting and use an additive and multiplicative Legendre polynomial, and for some cases, a trigonometric series to correct the continuum. 

In the fitting process, {\sc pPXF} determines the light-weighted contributions (weights) of each stellar population template. 
From the weightages of different populations of stars as found from the {\sc pPXF} fit, we found the age and the metallicity of the sources. We further determined the stellar mass of the sources by measuring the $ i$-band apparent magnitude, which in turn yields the luminosity and the mass-to-light (M/L) ratio in the corresponding band using {\sc pPXF}. We also derive the stellar mass by considering the photometry, i.e.,  the apparent magnitude of all the five SDSS bands ($u,\thinspace g, \thinspace r, \thinspace i, \thinspace \& \thinspace z$). Then the photometric measurements are converted to flux and included in the fit, ensuring that the stellar population model is normalised to the galaxy’s total observed light across a broad wavelength range. The best-fit template weights are then used to sum the stellar mass contributions from each population bin, scaled by the galaxy’s luminosity distance and observed flux. This approach also yields an estimate of the stellar mass, constrained by both the spectrum and photometry. The stellar masses obtained in both methods are very similar for all sources in our sample. We choose to use the stellar mass found in the former method  i.e., using $i$-band magnitude, for the rest of our analysis. 

To assess the uncertainty of the derived parameters, we implemented a bootstrap resampling procedure. Synthetic spectra are generated by perturbing the best-fit model with resampled residuals (wild bootstrap), and the {\sc pPXF} fitting is repeated for 50 realisations. The standard deviation of the distribution of recovered ages, metallicities, and stellar masses provides error estimates.

We fitted the spectra of all 1040 sources with SNR > 10 and manually cross-checked the quality of fits for the 915 sources previously flagged as good quality. In this second case, i.e., fitting the spectra without masking the gas emission lines, we confirmed that all 915 sources also exhibit good quality fits. These 915 sources therefore constitute our final sample used in further analysis, and all the results presented in the subsequent sections of this paper are based on this sample. Example of a good-quality fit from this second case is shown in Figure \ref{fig:ppxf_fit_stellar_gas}. Among the 915 sources, we identified 341 dual systems, i.e., systems in which both nuclei have SNR $>$ 10 and good-quality fits. In addition, we found 8 triplet systems, where all three nuclei satisfy these criteria. No quadrupole system has been identified to satisfy these criteria. It is to be noted that from our 915 finalised sample, if one of the sources in a dual or triple or quadrupole system does not exhibit a good quality fit, then we discard that system to study the properties of the dual, triple or quadrupole systems. 
Our subsequent analysis of dual-nuclei systems is therefore based on the 341 dual systems, derived from the original GOTHIC sample of 682 dual/multiple sources.

\section{Results}
\label{sec:results}
\subsection{Consistency check}
\label{subsec:consistency_check}
As we determine the stellar kinematics following the procedure mentioned in subsection \ref{subsec:stellar_kin}, we take the stellar dispersion velocity ($\sigma_{\star}$) from our analysis and compare it with the stellar velocity dispersion measured in SDSS spectral fitting. The left column of Figure \ref{fig:consitency_check} shows the comparison of stellar velocity dispersion between our analysis and that from SDSS. From these figures, we can see that the stellar dispersion velocity from our analysis is consistent with that from SDSS for most of the sources. However, for some sources, stellar velocity dispersion from SDSS is provided as equal to zero, but from our analysis using \ppxf  we obtain an acceptable value of the velocity dispersion. This may happen in the case of SDSS due to the limitations in template matching, low signal-to-noise ratios, or when the intrinsic velocity dispersion is close to the instrumental resolution. In these conditions, the automated SDSS fitting procedure cannot reliably distinguish broadening of the spectra and hence the stellar velocity dispersion. In contrast, \ppxf's flexible approach to template selection and its advanced fitting algorithms allow it to extract meaningful velocity dispersions even in these challenging cases, providing reliable measurements where the SDSS pipeline fails \citep{sdss_spec_fit_bernardi_2003, sdss_spectra_fit_Bolton_2012}.
Also, for one of the sources, we find the stellar velocity dispersion is too large in comparison to what we get from our analysis.
This issue may arise because the velocity dispersion measurements in SDSS spectra are based on template spectra convolved up to a maximum sigma of 420 kms. Therefore, stellar velocity dispersion values exceeding 420 kms are considered unreliable and should be discarded, as noted in the SDSS documentation.

As another cross-check, we compare the stellar masses derived from our analysis, as described in Section \ref{subsec:gas_lines}, with those obtained from FIREFLY \citep{Firefly_wlikinson_2017}. The right-hand panels of Figure \ref{fig:consitency_check} present this comparison. We find that, for almost all sources, FIREFLY systematically yields lower stellar masses than those obtained using \ppxf. This discrepancy is not new, while analysing the spatially resolved stellar population properties of nearly 10,010 nearby galaxies from  MaNGA \citep{Manga2015} survey, \citet{firefly_mass_diff_Neumann2022} (The MaNGA FIREFLY Value-Added-Catalogue) reported stellar masses that are $\sim 0.3$ dex higher than those derived using FIREFLY.

In our study, the median stellar mass of all sources from our analysis is $\log_{10}(M_\star/M_\odot) = 11.18\pm0.59$, whereas the median stellar mass of all the gothic sources from FIREFLY is 10.60, implying that our estimates are higher by $\sim 0.58$ dex compared to FIREFLY. As mentioned earlier, 32 sources are common between the MaNGA-FIREFLY \citep{firefly_mass_diff_Neumann2022} catalogue and our finalised sample. However, when considering all sources in the MaNGA-FIREFLY catalogue, the average stellar mass is  $\log_{10}(M_\star/M_\odot)$ = $10.91\pm0.44$.

These comparisons indicate a systematic offset between stellar masses derived from our \ppxf-based analysis and those obtained from FIREFLY, with FIREFLY yielding lower values by $\sim 0.58$ dex. However, the stellar masses from FIREFLY show a clear positive correlation with those derived in this work. This suggests that the observed difference primarily arises from methodological differences in the fitting approaches rather than random statistical uncertainties. FIREFLY determines the best-fitting model through an iterative process and can be more strongly influenced by the youngest, most luminous stellar populations, which outshine older, less luminous but more massive stars—an effect often referred to as the \say{iceberg effect} \citep{Firefly_wlikinson_2017}. In contrast, {\ppxf} employs regularization to recover a smoother star formation history, allowing the inclusion of underlying older stellar populations that contribute significantly to the total mass. This naturally can lead to higher stellar mass estimates. 
Besides that, this difference is consistent with previous MaNGA-based studies \citep{firefly_mass_diff_Neumann2022}. While the limited number of matched sources and the small overlap with other SDSS value-added catalogues prevent a robust comparison of absolute stellar masses for our sample, our average stellar mass remains broadly consistent with that of the overall MaNGA-FIREFLY \citep{firefly_mass_diff_Neumann2022} population.

\begin{figure}
    \centering
    \includegraphics[width=1\linewidth]{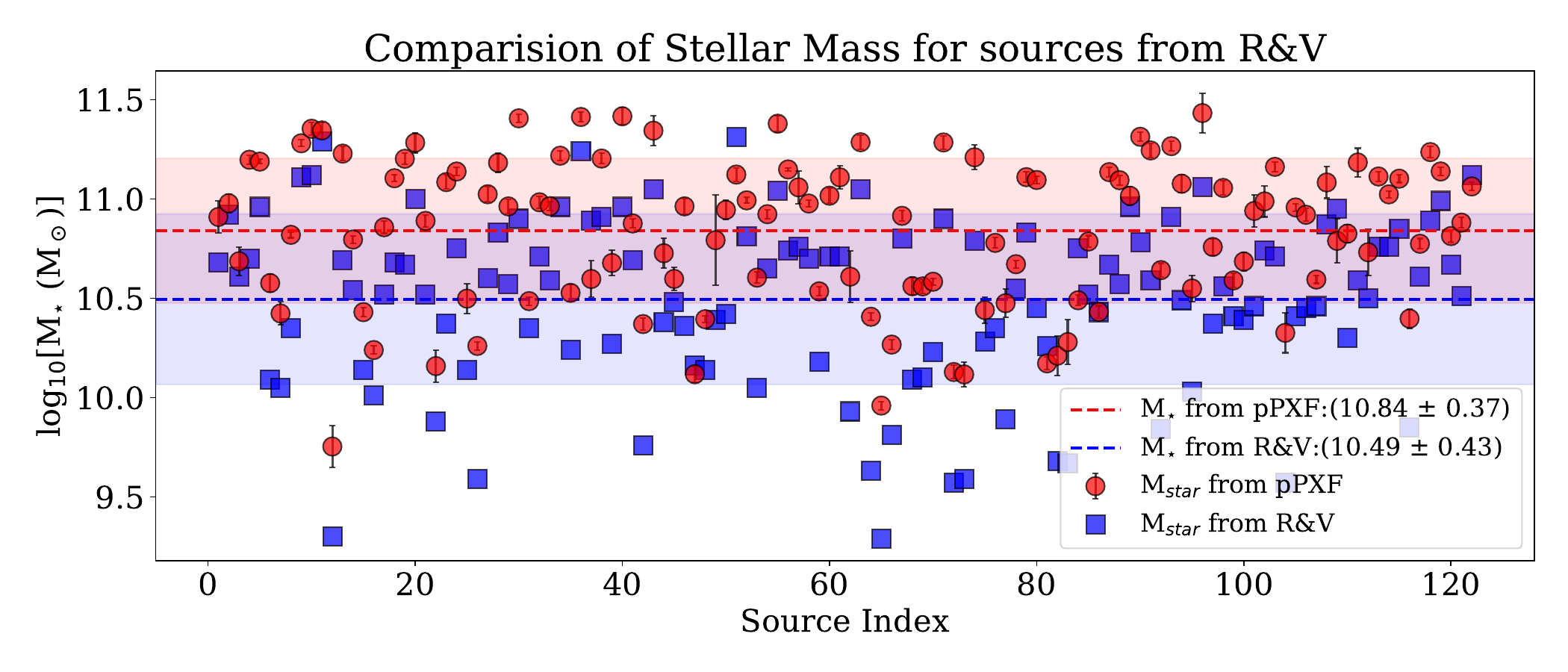}
    \caption{Comparison of stellar masses for the 123 sources from \citet{mbh_mstar_Reines2015}, derived using our \ppxf-based method, and those reported in their study.}
    \label{fig:RV_mass_comp}
\end{figure}

Because of the above-mentioned limitations in comparing stellar masses for our sample, we carried out an additional consistency check to further assess the reliability of our process of stellar mass estimation. We retrieved the SDSS spectra of 262 sources from \citet{mbh_mstar_Reines2015} and derived their stellar masses using our \ppxf-based methodology. From these, we chose 123 single-nucleus sources with input spectral SNR $>10$ and good fitting quality (qflag = 1). For this subsample, we compared the stellar masses obtained from our method with those reported by \citet{mbh_mstar_Reines2015}. It is worth noting that the stellar masses in their study were primarily estimated using the $i$-band mass-to-light ratio. Figure~\ref{fig:RV_mass_comp} presents the corresponding comparison. The mean stellar mass of this sample, derived using our \ppxf-based method, is $10.84 \pm 0.37$, while the average mass derived from the masses reported by \citet{mbh_mstar_Reines2015} is $10.49 \pm 0.43$. The comparison shows that although our method finds masses a factor $\sim 2$ higher than \citet{mbh_mstar_Reines2015}'s method for the same single-nuclei galaxies, the difference is in the scatter.

We therefore conclude that, although absolute stellar masses may differ from FIREFLY-based estimates, our stellar mass measurements provide a consistent basis for internal comparisons and for studying different properties within the sample.

\subsection{Emission Line-Based Classification}
By fitting the spectra as described in subsection~\ref{subsec:gas_lines}, we measured the fluxes of the available gas emission lines as given by \ppxf. We then classified the sources for which all four lines, H$\alpha$, H$\beta$, [OIII]$\lambda5007$, and [NII]$\lambda6583$ were detected, following the Baldwin, Phillips, and Terlevich (BPT) classification scheme \citep{bladwin1980}. The sources were categorised as star-forming (SF), AGN–LINER, AGN–Seyfert, or Composite. Out of the 915 finalised sources in our sample, 53 were classified as star-forming, 142 as AGN–LINER, 79 as AGN–Seyfert, and 136 as composite. Hence, the classification was possible for nearly 45\%, i.e., 410 sources and  55\% of the sample, corresponding to 505 sources, could not be classified due to the non-detection of one or more of the four required lines. The corresponding BPT diagram for the classified sources is shown in Figure~\ref{fig:bpt}. 

It is to be noted that \citet{gothic_2023} also studied the classification of sources based on the BPT diagram for our initial sample of  1393 galaxies using fluxes from the SDSS database. From the SDSS database, line fluxes for all four lines were recovered for 1098 sources. This larger number of sources compared to our  \ppxf-based measurements is mostly because SDSS fits emission lines directly to the observed spectra, providing fluxes even for spectra with low signal-to-noise or weak emission features, regardless of the quality of the stellar continuum subtraction. In contrast, \ppxf first performs a careful fit and subtraction of the stellar continuum, and only reports reliable emission line fluxes when both the continuum fit and the emission line detection are robust. As a result, our \ppxf-based classification is more conservative, including only sources with high-quality fits and significant emission line detections, while the SDSS catalogue is more inclusive but may contain less reliable measurements for faint or noisy lines from the spectra. However, a more accurate and precise way of finding the gas emission lines that can recover the line fluxes even for lines with lower SNR can be done using {\ppxf} and GANDALF: Gas AND Absorption Line Fitting \citep{Gandolf_software2017}, following the procedure by \citet{portsmouth_thomas_2013}. Currently, we keep it as a future aspect of this study on GOTHIC sources and will be explored in detail in the upcoming papers of the GOTHIC series.

\begin{figure}
    \centering
    \includegraphics[width=\linewidth]{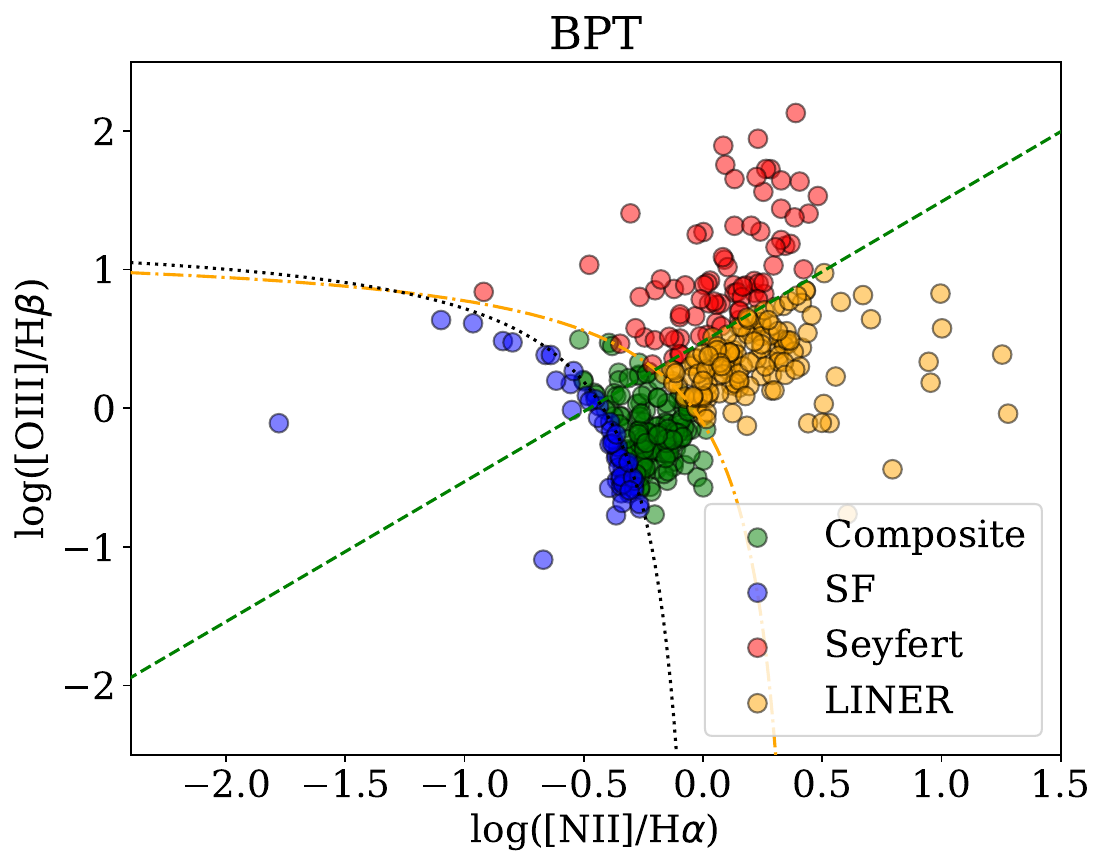}
    \caption{The BPT diagram of the 410 sources for which all four lines were detected. The black dotted line represents the lower limit for AGNs (both LINERs and Seyferts) \citep{Kauffmann2003}. The yellow dash-dotted curve represents the upper limit for starburst \citep{Kewley2006}. The dashed green line separates the LINERs from Seyferts \citep{Fernandes2010}. Hence, the blue, green, red, and yellow data points represent the Star-forming, Composites, AGN-Seyferts, and AGN-LINERs, respectively.}
    \label{fig:bpt}
\end{figure}

\subsection{Distribution of Source Properties}
\label{subsec:dist_sour_prop}

\begin{figure*}
    \centering
        \begin{subfigure}[b]{0.32\textwidth}
         \centering
         \includegraphics[width=\textwidth]{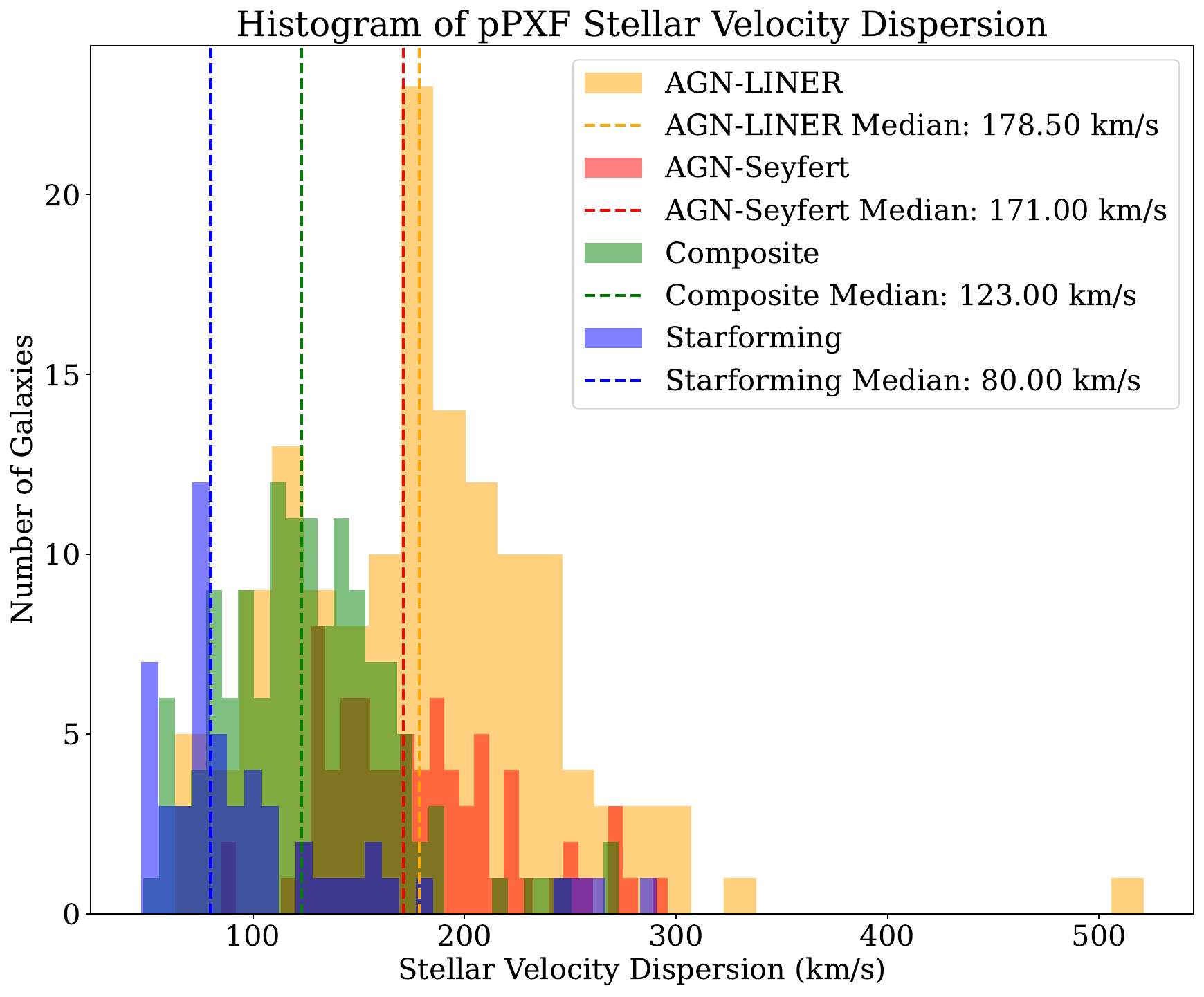}
         \caption{}
         \label{fig:sigma_dist}
        \end{subfigure}
        \hfill
        \begin{subfigure}[b]{0.32\textwidth}
         \centering
         \includegraphics[width=\textwidth]{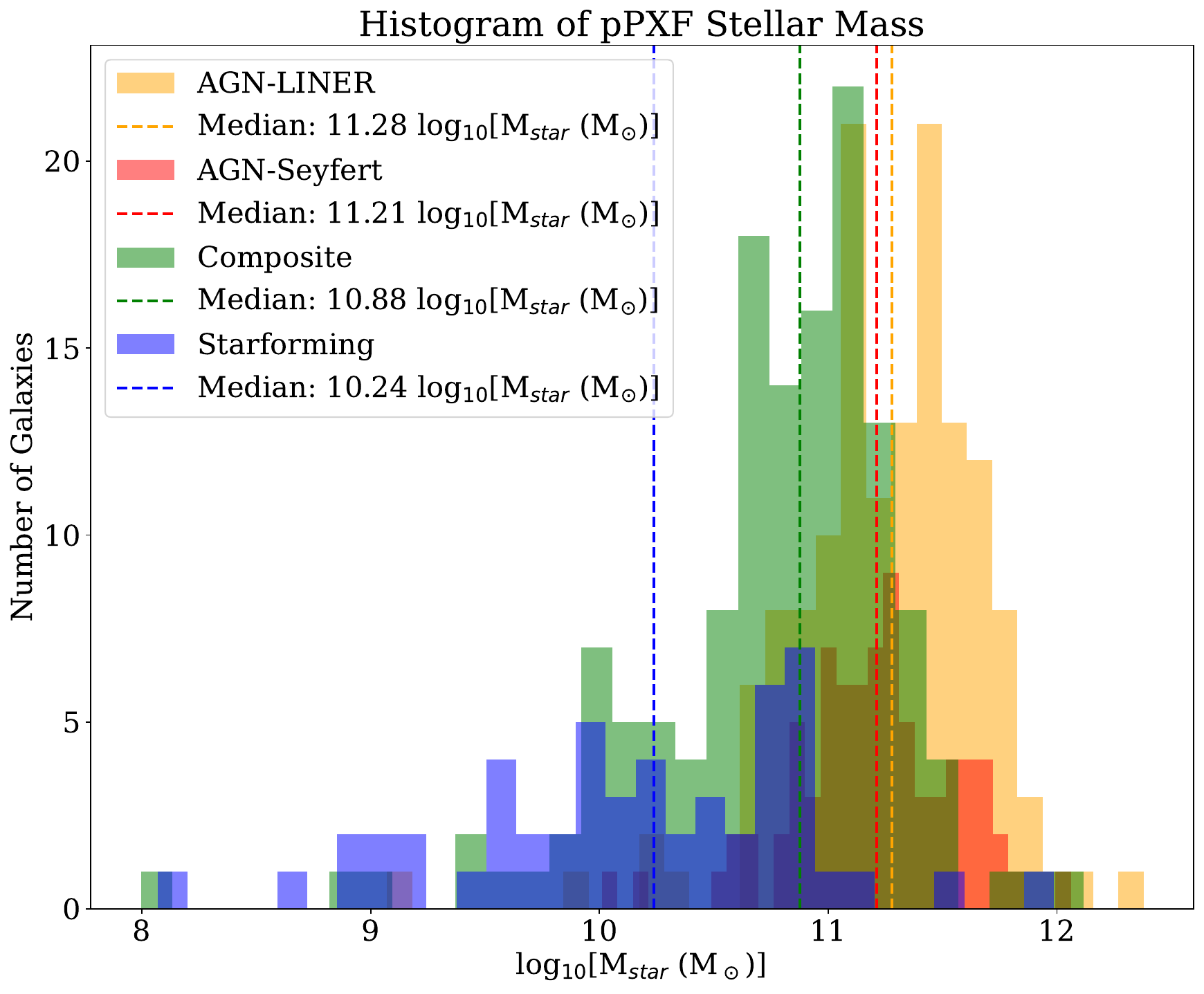}
         \caption{}
         \label{fig:mass_dist}
        \end{subfigure}
        \hfill
        \begin{subfigure}[b]{0.32\textwidth}
         \centering
         \includegraphics[width=\textwidth]{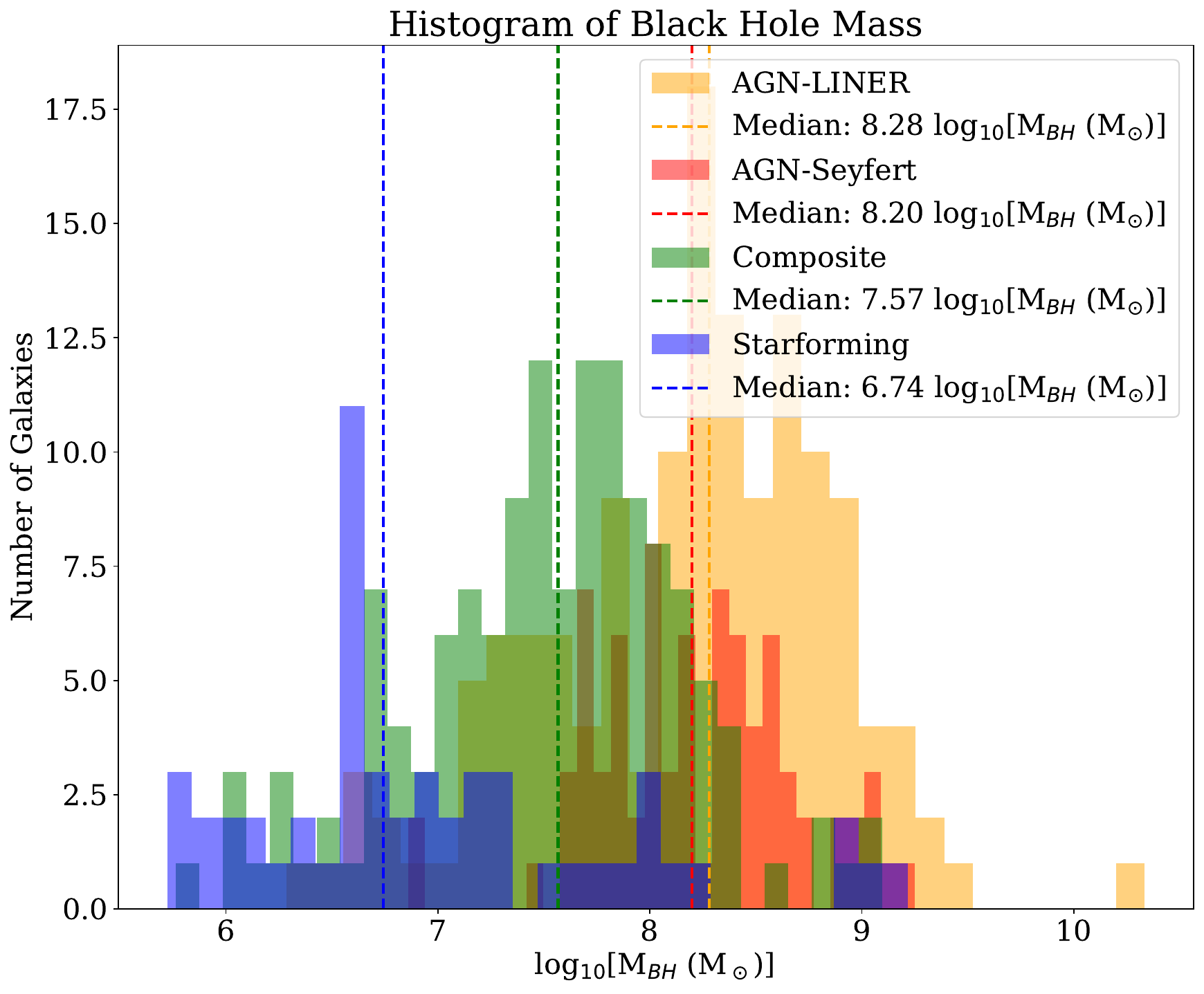}
         \caption{}
         \label{fig:mbh_dist}
        \end{subfigure}
        \hfill
        \begin{subfigure}[b]{0.32\textwidth}
         \centering
         \includegraphics[width=\textwidth]{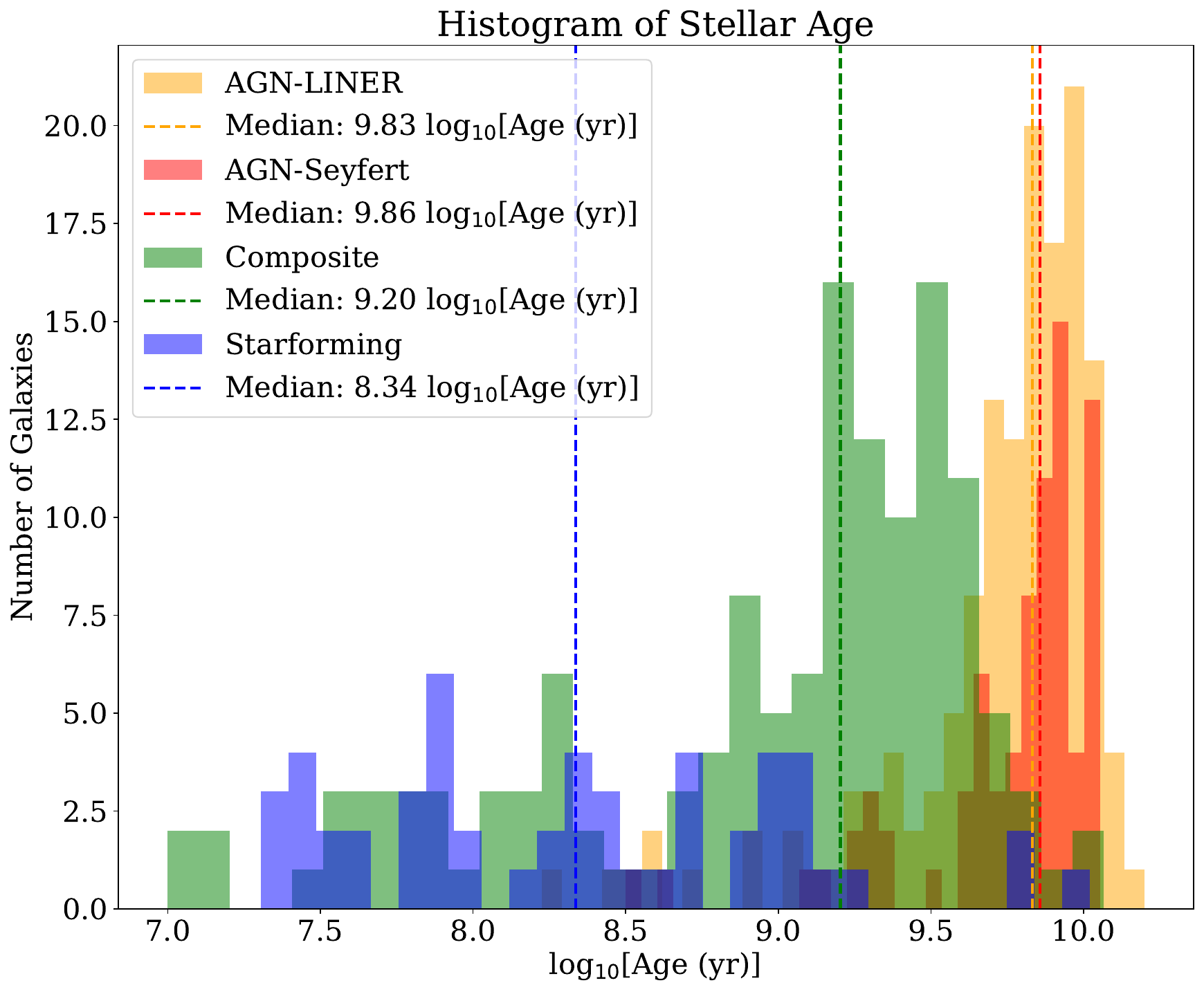}
         \caption{}
         \label{fig:age_dist}
        \end{subfigure}
        \hfill
        \begin{subfigure}[b]{0.32\textwidth}
         \centering
         \includegraphics[width=\textwidth]{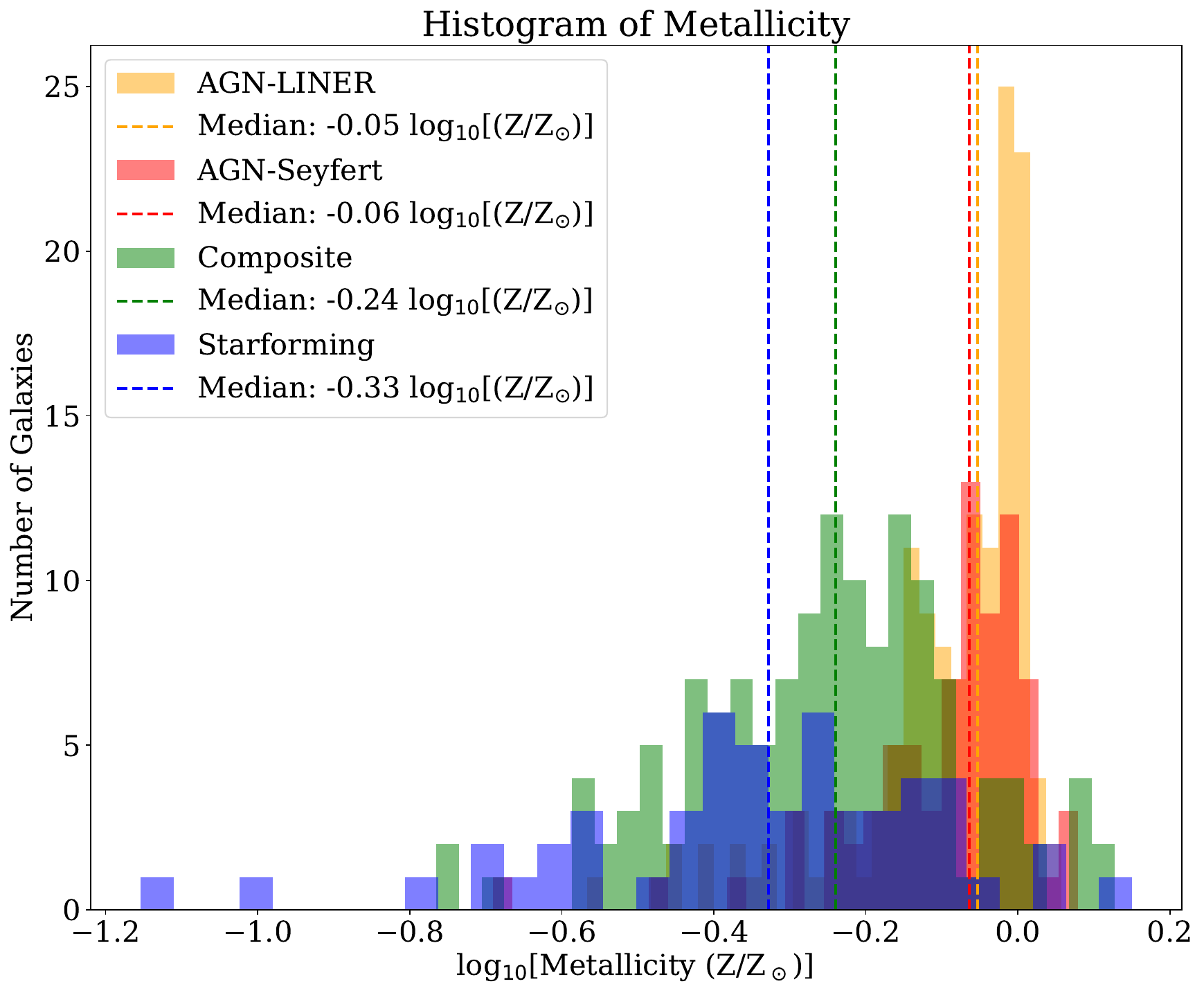}
         \caption{}
         \label{fig:metal_dist}
        \end{subfigure}
        \hfill
        \begin{subfigure}[b]{0.32\textwidth}
         \centering
         \includegraphics[width=\textwidth]{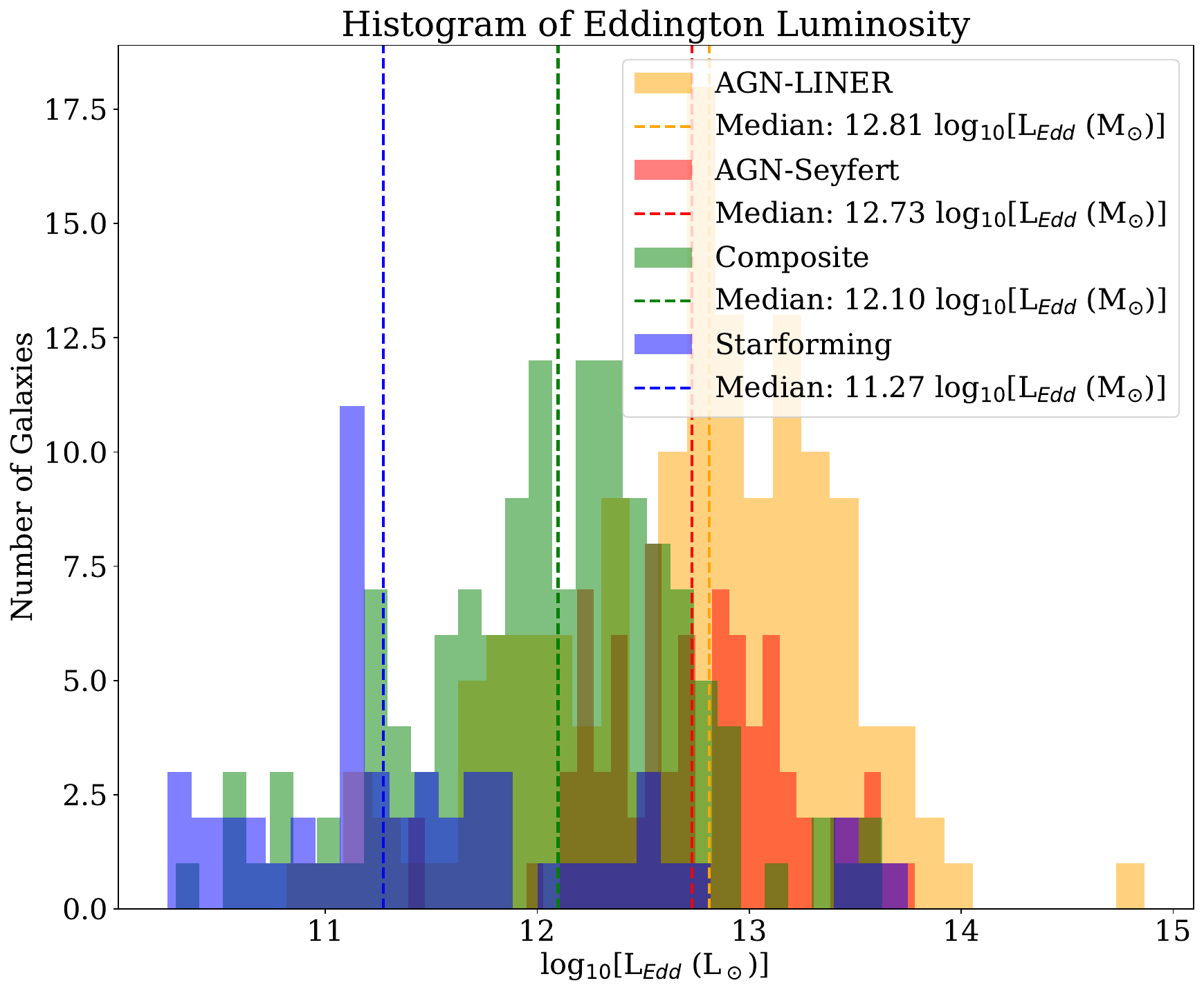}
         \caption{}
         \label{fig:edd_dist}
        \end{subfigure}
        
    \caption{Distributions of nuclear properties of 410 classified sources,  including stellar velocity dispersion, stellar mass, black hole mass, stellar age, metallicity,  and Eddington luminosity. The  orange, red, green, and blue colors represent the distributions of the AGN-LINER, Seyfert, Composite, and star-forming sources, respectively. The dashed lines indicate the median values for each class.}
    \label{fig:all_prop_dist}
\end{figure*}

\begin{table*}
    \centering
    \begin{tabular}{ccccc}
    \hline
    \hline
     Types:  & Star-forming & Composite & AGN-LINER & AGN-Seyfert \\
    \hline
    \hline
    Number of sources in each type: & 53 & 136 & 142 & 79 \\
    \hline
    Medians: & & & & \\
    Stellar dispersion velocity ($\text{km}\thinspace \text{s}^{-1}$) &  80 & 123 & 171 & 208 \\
    Stellar Mass ($\text{log}_{10}[\text{M}_{star}(\text{M}_{\odot})]$) & 10.24 & 10.88 & 11.21 & 11.28 \\
    Black hole Mass ($\text{log}_{10}[\text{M}_{BH}(\text{M}_{\odot})]$) & 6.74 & 7.57 & 8.20 & 8.28 \\
    Stellar age ($\text{log}_{10}[\text{Age}(\text{yr})]$) & 8.34 & 9.20 & 9.86 & 9.83 \\
    Metallicity ($\text{log}_{10}[z/z_{\odot}]$) & -0.33 & -0.24 & -0.06 & -0.05  \\
    M/L ($r$-band) & 1.73 & 3.00 & 6.74 & 6.48 \\
    Eddington Luminosity ($\text{log}_{10}[L_{Edd}(L_{\odot})]$) & 11.27 & 12.10 & 12.73 & 12.81 \\
    \hline
    \hline
    \end{tabular}
    \caption{Number of sources and the median values of their various properties for each category identified in our sample.}
    \label{tab:median}
\end{table*}

We examined the distribution of several physical and spectral properties of all 915 sources. Figure~\ref{fig:sigma_dist} presents the distribution of stellar velocity dispersion for our sample. The overall median velocity dispersion is 180 km s$^{-1}$. Star-forming sources lie at the lower end of the distribution, with a median of 80 km s$^{-1}$. Composite sources show intermediate values, with a median of 123 km s$^{-1}$, higher than the star-forming systems but lower than the AGN. The AGN exhibit the highest dispersions, with a median of 171 km s$^{-1}$ for Seyfert and 178.5 km s$^{-1}$ for LINERS and values extending up to $\sim$300 km s$^{-1}$. These distributions of dispersion velocity for different types of sources are in good agreement with previous studies \citep[][, etc.]{Sheth2003, portsmouth_thomas_2013, sigma_agn_Koss2022}. 

The stellar mass distribution (Figure~\ref{fig:mass_dist}) also shows a similar trend. The median stellar mass of the full sample is $\log(M_\star/M_\odot)=11.26$. AGNs, including AGN-LINERs and Seyferts, exhibit comparable median stellar masses of $\log(M_\star/M_\odot)=11.28$ and $11.21$, respectively. In contrast, Composite and star-forming galaxies have systematically lower median stellar masses, with $\log(M_\star/M_\odot)=10.88$ and $10.24$, respectively.

Similar to the stellar mass and stellar dispersion velocity, the black hole masses (Figure~\ref {fig:mbh_dist}) exhibits systematic variations across different spectral classes. The black hole masses shown in Figure~\ref{fig:mbh_dist} are found using the M$_{BH}$ - $\sigma$ relation \citep{Kormendy_Ho_2013}: 
\begin{equation}
\log_{10}\left(\frac{M_{\rm BH}}{10^{9} M_\odot}\right)
= a + b \, \log_{10}\left(\frac{\sigma}{200 \,\mathrm{km\,s^{-1}}}\right)
\end{equation}

where  M$_{BH}$ is the mass of the central black hole of the galaxy, $\sigma$ is the stellar dispersion velocity, $a = -0.501 \pm 0.049$ and $b = 4.414 \pm 0.295$. The median black hole mass for the entire sample is $\log(M_{\rm BH}/M_\odot)=8.30$. Systems hosting AGN are characterized by relatively massive black holes, with AGN-LINERs and Seyferts showing median values of $\log(M_{\rm BH}/M_\odot)=8.28$ and $8.20$, respectively. In comparison, Composite galaxies and star-forming systems harbour significantly less massive black holes, with median masses of $\log(M_{\rm BH}/M_\odot)=7.57$ and $6.74$.

The stellar age distribution shows systematic differences across the spectral classes (Figure~\ref {fig:age_dist}). The median stellar age of the full sample, $\log_{10}(\mathrm{Age/yr})$, is 9.87. AGN-LINER and Seyfert nuclei exhibit similarly old stellar populations, with median ages of $\log_{10}(\mathrm{Age/yr})$=9.83 and 9.86, respectively. Composite nuclei also host relatively old stellar populations, with a median age of $\log_{10}(\mathrm{Age/yr})$=9.20, whereas star-forming nuclei are characterised by significantly younger stellar populations, with a median age of $\log_{10}(\mathrm{Age/yr})$=8.43.

The metallicity distribution also varies across the spectral classes (Figure~\ref {fig:metal_dist}). The median metallicity for the full sample is $\log(Z/Z_\odot) = -0.06$. AGN-LINER and Seyfert nuclei exhibit comparable metallicities, with median values of $\log(Z/Z_\odot) = -0.05$ and $-0.06$, respectively. Composite nuclei show moderately lower metallicities, with a median of $\log(Z/Z_\odot) = -0.24$, while star-forming nuclei are characterized by the lowest metallicities, with a median value of $\log(Z/Z_\odot) = -0.33$. This trend is consistent with the stellar age distribution, with older nuclei generally exhibiting higher metallicities, while younger, star-forming nuclei tend to be more metal-poor.

The distribution of the mass-to-light ratio at $r$-band also shows clear differences across the spectral classes. The median mass-to-light ratio for the full sample is 6.65. AGN-LINER and Seyfert nuclei exhibit similarly high mass-to-light ratios, with median values of 6.48 and 6.74, respectively. In contrast, Composite nuclei show substantially lower mass-to-light ratios, with a median of 3.00, while star-forming nuclei have the lowest values, with a median of 1.73. This behaviour is consistent with the stellar age and metallicity trends, as older and more metal-rich nuclei tend to have higher mass-to-light ratios, while younger, metal-poor nuclei exhibit lower values.

Similarly, the Eddington luminosity distribution also varies across the spectral classes (Figure~\ref {fig:edd_dist}).  Similarly, the Eddington luminosity \citep{Eddington1926} in Figure~\ref{fig:edd_dist} is found from black hole mass following \citet{Ledd_vietri_2008}:
\begin{equation}
\begin{aligned}
L_{\rm Edd} &= 1.3 \times 10^{38} \left(\frac{M_{\rm BH}}{M_\odot}\right)\,{\rm erg\,s^{-1}} \\
             &\simeq 3.4 \times 10^{4} \left(\frac{M_{\rm BH}}{M_\odot}\right) L_\odot
\end{aligned}
\end{equation}

where L$_{Edd}$ is the Eddington luminosity and M$_{BH}$ is the mass of the central black hole. The median Eddington luminosity for the full sample is $\log[L_{\rm Edd}(L_\odot)] = 12.83$. AGN-LINER and Seyfert nuclei show comparable Eddington luminosities, with median values of $\log[L_{\rm Edd}(L_\odot)] = 12.81$ and $12.73$, respectively. Composite nuclei exhibit lower Eddington luminosities, with a median of $\log[L_{\rm Edd}(L_\odot)] = 12.10$, while star-forming nuclei have the lowest values, with a median of $\log[L_{\rm Edd}(L_\odot)] = 11.27$.

Thus, in each case, star-forming galaxies occupy the lower end of the distribution. Composite galaxies typically show higher values than star-forming galaxies but lower values than AGN. The median values of various properties for different categories of sources are shown in Table \ref{tab:median}. 
The observed trend in the distribution of all the 410 classified sources are shown in Figure~\ref{fig:all_prop_dist} and can be understood in the context of galaxy evolution. Star-forming galaxies are generally less massive, disk-dominated systems with relatively young stellar populations and lower metallicities, reflecting their ongoing star formation and less enriched interstellar medium. Composite systems, which show contributions from both star formation and nuclear activity, occupy an intermediate regime in stellar mass, black hole mass, and stellar population properties, consistent with their transitional nature. AGN hosts, by contrast, are typically massive, bulge-dominated galaxies with older and more metal-rich stellar populations, deeper gravitational potentials, and more massive central black holes. Together, these trends are consistent with an evolutionary scenario in which galaxies grow in mass, quench star formation, and transition from star-forming to AGN-dominated systems. However, we note that a similar behaviour could also arise from mass-dependent effects, where less massive systems remain in their respective regimes without necessarily undergoing such evolution.

\subsection{Investigating interdependence of various properties in comparison to the single-nuclei systems}

\begin{figure*}
\centering

\begin{subfigure}[t]{0.7\textwidth}
    \centering
    \includegraphics[width=\textwidth]{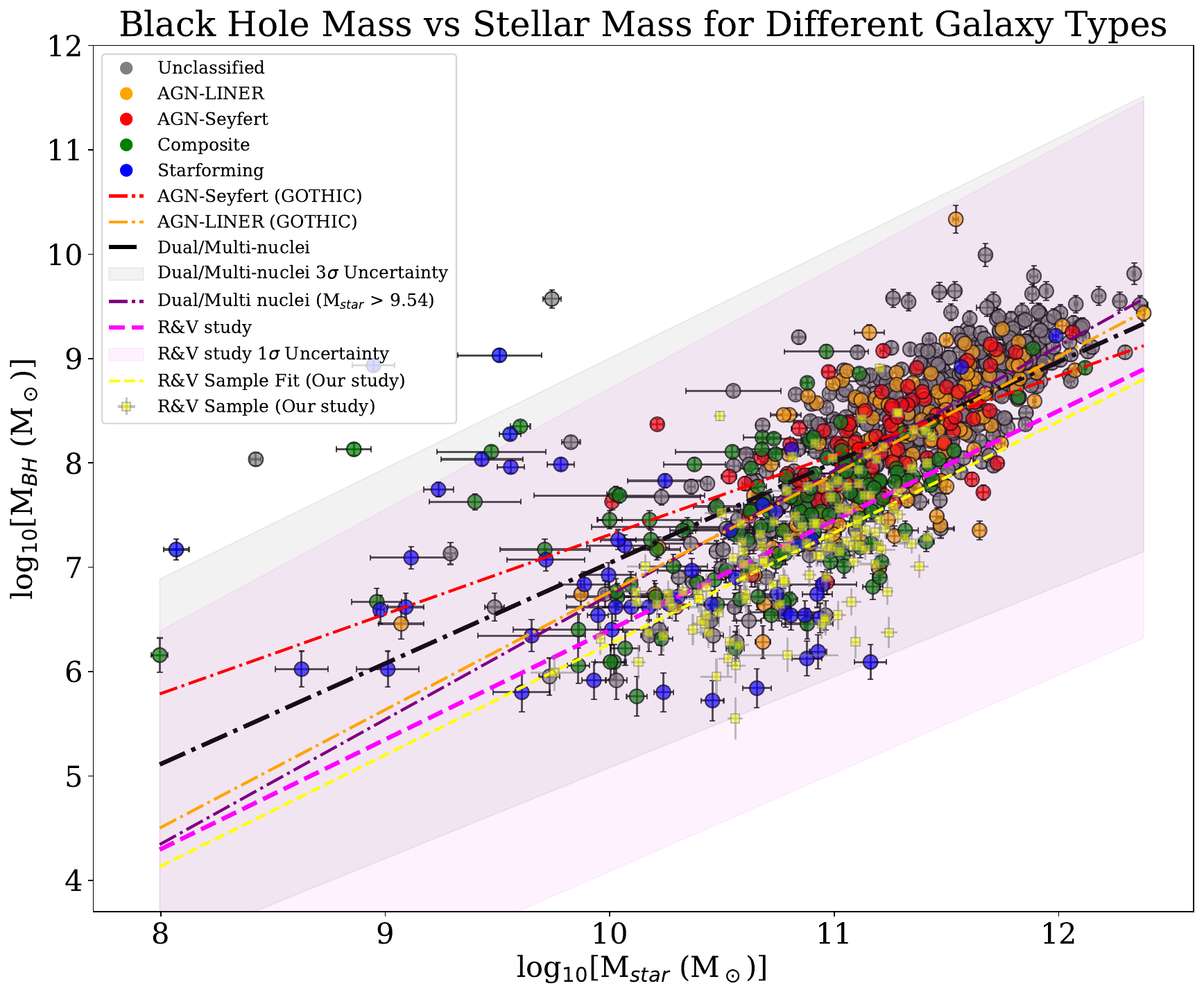}
    \caption{Black hole mass Vs of stellar mass. Blue, green, red, orange, and gray circles represent star-forming, composite, AGN--Seyfert, AGN--LINER, and unclassified nuclei, respectively, from the our sample. The black dash-dotted line shows the best-fit relation for all dual- and multi-nuclear systems in our sample, with the gray shaded region indicating the corresponding $3\sigma$ scatter. The red, orange, and violet dash-dotted lines denote the best-fit relations for AGN--Seyfert, AGN--LINER, and sources with $M_{\star} > 9.45$, respectively. For comparison, the magenta dashed line and shaded region represent the relation and its scatter reported by \citet{mbh_mstar_Reines2015} (R\&V). Yellow points correspond to selected 123 single-nucleus sources from R\&V for which $M_{\star}$ and $M_{\mathrm{BH}}$ have been re-estimated using our {\ppxf}-based method, and the yellow solid line shows the best-fit relation to these recalculated values.}
    \label{fig:corr_mbh_mstar}
\end{subfigure}

\vspace{0.6em}

\begin{subfigure}[t]{0.32\textwidth}
    \centering
    \includegraphics[width=\textwidth]{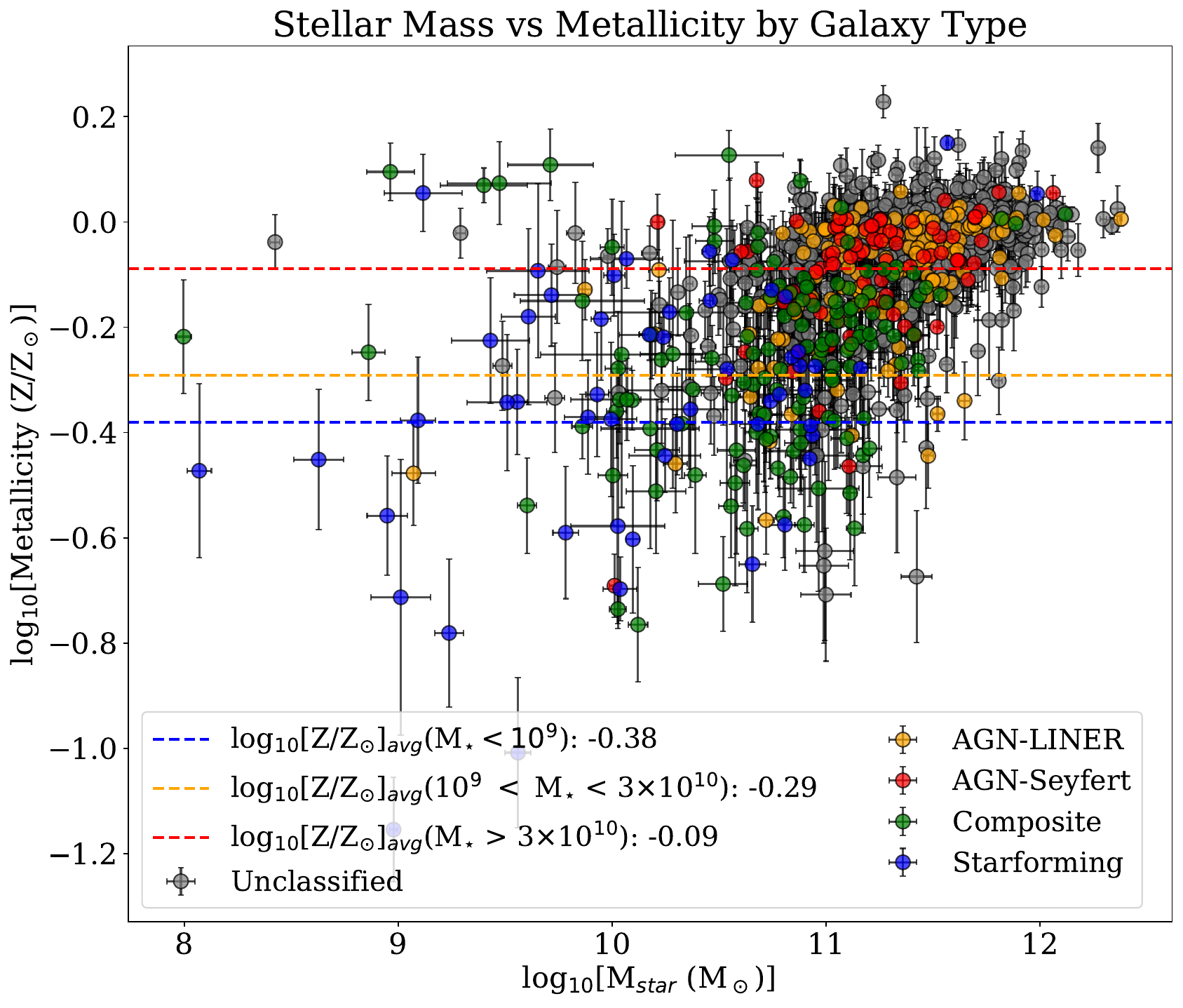}
    \caption{Metallicity Vs stellar mass. Blue, green, red, orange, and gray circles represent star-forming, composite, AGN--Seyfert, AGN--LINER, and unclassified nuclei, respectively, from the our sample. Blue, yellow and red dashed lines respectively reprents the average metalicity for $M_\star<10^9$, $10^9<M_\star<3\times10^{10}$ and $M_\star > 3\times 10^{10}$.}
    \label{fig:corr_metal_mstar}
\end{subfigure}
\hfill
\begin{subfigure}[t]{0.32\textwidth}
    \centering
    \includegraphics[width=\textwidth]{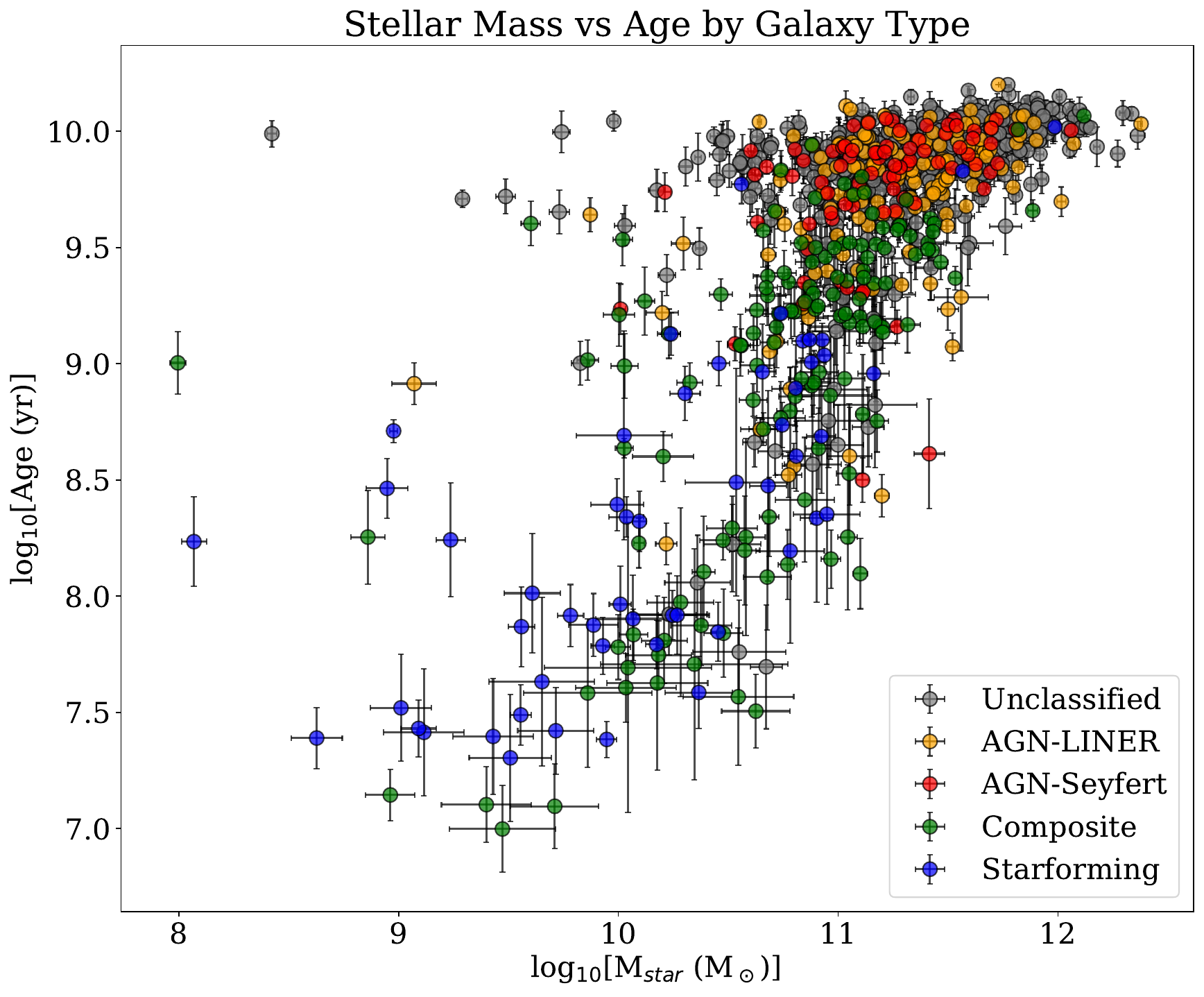}
    \caption{Age Vs stellar mass. Blue, green, red, orange, and gray circles represent star-forming, composite, AGN--Seyfert, AGN--LINER, and unclassified nuclei, respectively, from the our sample.}
    \label{fig:corr_age_mstar}
\end{subfigure}
\hfill
\begin{subfigure}[t]{0.32\textwidth}
    \centering
    \includegraphics[width=\textwidth]{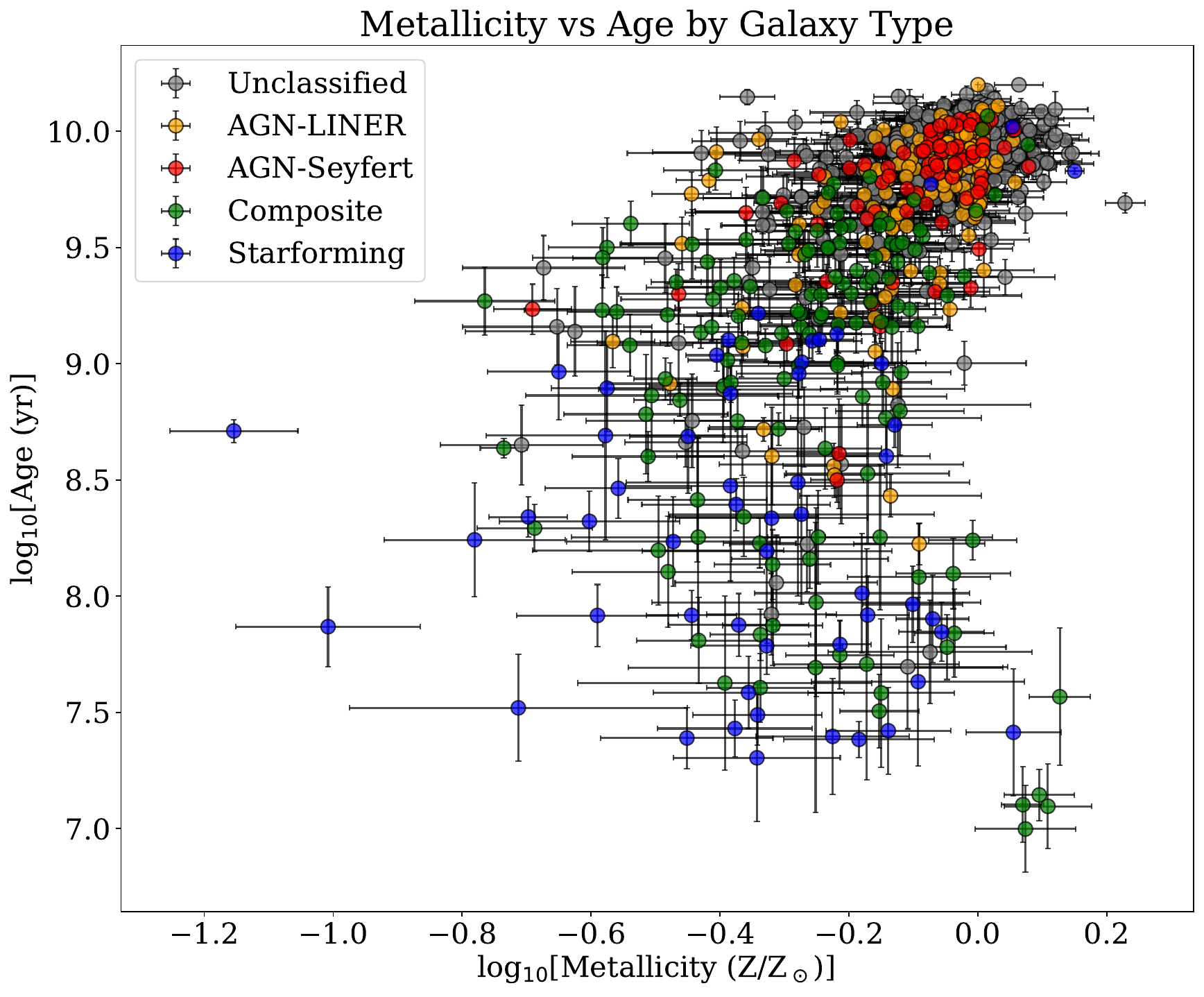}
    \caption{Age Vs Metallicity. Blue, green, red, orange, and gray circles represent star-forming, composite, AGN--Seyfert, AGN--LINER, and unclassified nuclei, respectively, from the our sample.}
    \label{fig:corr_age_metal}
\end{subfigure}

\caption{
Correlation between different physical properties of the sources in our sample.
}
\label{fig:comp_all}

\end{figure*}

We further explored correlations between stellar mass, velocity dispersion, age, metallicity, and mass-to-light ratio, both for the final 915 sources from our sample, and also according to the different nuclei types. It is important to note that in some multi-nuclei systems, one or more companions were removed from the finalised sources due to poor-quality fits (qflag = 3 or 4). However, the remaining companions in those systems are still valid members of the multi-nuclei system. Therefore, including these systems along with those in which all companions have reliable fits, preserves the overall characteristics of multi-nuclei systems. Studying the properties of all 915 sources thus reflects the intrinsic behaviour of galaxy nuclei in closely merging galaxies. By comparing the properties these systems with those of single nuclei systems, we can assess the effect of the merging process on galaxy properties. In the following subsections, we explore these aspects in detail.

\subsubsection{Black hole mass Vs Stellar mass }
\label{subsubsec:mbh_mstar}
It is well observed in the literature that black hole mass and stellar mass exhibit a positive correlation \citep[e.g., see][, etc.]{mbh_mstar3_Shu2020, mbh_mstar2Zhu2021, mbh_mstar1_Li2023}. Our results are consistent with this trend. As shown in Figure~\ref{fig:corr_mbh_mstar}, the stellar mass generally increases with black hole mass across the sample, with the correlation being particularly strong for Composite nuclei and AGNs. In contrast, star-forming galaxies display a weaker, more scattered relation compared to the Composite and AGN populations. For the full set of sources in our sample, ranging from star-forming galaxies to AGNs, we obtain a slope of $0.96 \pm 0.03$ and an intercept of $-2.6 \pm 0.3$. When examined by AGN subtype, the AGN-Seyfert population yields a slope of $0.76 \pm 0.12$ with an intercept of $-0.300 \pm 1.396$, while AGN-LINERs show a steeper slope of $1.13 \pm 0.09$ and an intercept of $-4.519 \pm 1.098$. Furthermore, restricting the sample to galaxies with stellar masses $M_{\star} > 9.54$, as suggested by \citet{mass_metal_dwarf_Li2023} for the $M_{\rm BH}$--$M_\star$ relation, yields a slope of $1.19 \pm 0.03$ and an intercept of $-5.2 \pm 0.4$.

We further compare our results which is based on dual nuclei systems with another sample of AGN from \citet{mbh_mstar_Reines2015}.  This sample consisted of 242 broad-line AGNs from SDSS, and reports a slope of $1.05 \pm 0.11$ and an intercept of $-4.1 \pm 1.2$ for the $M_{\rm BH}$--$M_\star$ relation. \citet{mbh_mstar_Reines2015} do not distinguish between single- and multi-nucleus systems; therefore, their reported relation represents an average behaviour. Their sample also consists mostly of galaxies with $\log(M_\star/M_\odot) > 9$ (see Figure 5 of \citet{mbh_mstar_Reines2015}).  Hence, as also mentioned earlier in sub-section~\ref{subsec:consistency_check}, we selected 123 sources from the sample of \citet{mbh_mstar_Reines2015} with spectra having SNR $> 10$, good {\ppxf} fits (qflag = 1), and visually confirmed single-nucleus morphology. We then derived the stellar masses and black hole masses of these systems using the same \ppxf-based method applied to our sample. The best-fitting $M_{\rm BH}$--$M_\star$ relation for these sources yields a slope of $1.06 \pm 0.12$ and an intercept of $-4.38 \pm 1.36$. These values are in excellent agreement with those reported by \citet{mbh_mstar_Reines2015}. This agreement has two important implications: first, it validates our methodology for deriving stellar and black hole masses, despite the use of a different approach compared to \citet{mbh_mstar_Reines2015} and second, the deviation observed in the slope of the $M_{\rm BH}$--$M_\star$ relation for our dual/multi-nucleus systems, relative to this single-nucleus sample indicates that black hole–stellar mass relation may evolve differently in merging galaxies compared to isolated, single-nucleus galaxies, and is due to merger-driven growth. 
A clear evidence for this is that nearly all the dual AGN lie above the magenta or yellow dashed line in figure \ref{fig:corr_mbh_mstar}, which indicates that, for similar stellar masses, dual AGNs have larger SMBH masses. This shows that SMBH mass grows during the merging process and not just due to SMBH coalescence. It has long been suggested in the literature that gas inflows and nuclear star formation results in the growth of SMBHs during galaxy interactions \citep{Hopkins2006, Ellision2013_instab_merger3, Satyapal2014}. This is especially true in gas rich merging galaxies and where the nuclear starbursts result in bulge growth and mass accretion onto the SMBHs, often triggering dual AGN  \citep{Hopkins2006, Hopkins2008, Satyapal2014}. This is the first time that it has been so clearly shown. 

A large fraction of the composite nuclei also lie above the dashed line, indicating that they too are accreting mass and have SMBH masses larger than single AGN. However, they have slightly more scatter, which is probably because these nuclei generally host AGN and star formation activity. This can lead to some scatter in the SMBH mass estimates.  The star forming nuclei have the largest scatter and are in the lower mass galaxies. 

Although SMBH and stellar mass growth remain closely coupled across a wide range of source types, the steeper slope observed for AGN-LINERs and for galaxies above the stellar-mass threshold implies more efficient black hole growth relative to stellar mass in massive and likely more evolved systems. A comparatively shallower slope for AGN-Seyferts suggests black hole growth increases more slowly with stellar mass than in other AGN classes. 

In addition, this study identifies six sources that lie beyond the $2\sigma$ uncertainty of the $M_{\rm BH}$--$M_\star$ relation derived in this work. These outliers are characterised by relatively low stellar masses, $\log[M_{\star}/M_{\odot}] \sim 8.9$, yet host unusually massive black holes with $\log[M_{\rm BH}/M_{\odot}] \sim 8.8$, i.e., as massive as the host galaxies.



\subsubsection{Metallicity Vs Stellar mass}
In case of the metallicity with stellar mass, we find a general trend of increase in metallicity with the increasing stellar mass. However, the rate of increase varies for different types and different mass ranges of the sources. There is significant scatter in the mass-metallicity relation, especially at lower and intermediate stellar masses ($M_\star < 10^{9}\,M_\odot$). The increase in metallicity is rapid at intermediate masses and flattens at higher masses ($M_\star > 3\times10^{10}\,M_\odot$). The similar scenario has also been observed in the studies by \citep{mzr_Gallazzi2005} in these different mass bins. When dividing by stellar mass, we obtain: low-mass galaxies ($M_\star < 10^9\,M_\odot$) have average metallicity $\log_{10}[Z/Z_\odot]_{\rm avg} = -0.38 \pm 0.36$, intermediate-mass galaxies ($10^9 \leq M_\star \leq 3\times10^{10}\,M_\odot$) have $\log_{10}[Z/Z_\odot]_{\rm avg}$ = $-0.29 \pm 0.22$, and high-mass galaxies ($M_\star > 3\times10^{10}\,M_\odot$) have $\log_{10}[Z/Z_\odot]_{\rm avg}$ = $-0.09 \pm 0.14$. For lower mass galaxies, we get higher values in metallicity in comparison to \citet{mzr_Gallazzi2005} ($\log_{10}[Z/Z_\odot]_{\rm avg} \sim -0.6$) and for galaxies with higher mass, we get lower value in average metallicity in comparison to \citet{mzr_Gallazzi2005} ($\log_{10}[Z/Z_\odot]_{\rm avg} \sim 0.15$). We also measured the average metallicity for different galaxy types in our sample. For galaxy types, we find: AGN-LINER galaxies have an average metallicity of  $-0.10 \pm 0.12$, AGN-Seyfert $-0.10 \pm 0.12$, Composite $-0.25 \pm 0.17$, and Starforming $-0.34 \pm 0.25$.
These results suggest that multi-nuclei systems broadly follow the expected mass–metallicity trend but show systematic offsets compared to the classical relation from \citet{mzr_Gallazzi2005}. The higher metallicities at low stellar masses and lower metallicities at high stellar masses may reflect the impact of interactions or mergers in multi-nucleus systems, which can redistribute gas, trigger inflows, or dilute central metallicities. This indicates that the chemical evolution of multi-nuclei galaxies may be influenced by their dynamical state, leading to deviations from the standard mass–metallicity relation seen in isolated systems.


\subsubsection{Age Vs Stellar mass}
For galaxy ages, we find that star-forming and Composite galaxies exhibit a rapid increase in age with increasing stellar mass, whereas AGNs show a more gradual rise in age as stellar mass increases (see Figure~\ref{fig:corr_age_mstar}). The rapid increase in age with stellar mass for star-forming and Composite galaxies suggests that more massive galaxies in these categories host older stellar populations, consistent with the idea of accelerated or earlier star-formation histories (\say{downsizing}, \citet{mzr_Gallazzi2005}). In contrast, the slower age increase in AGNs implies that AGN hosts may have more diverse or extended star-formation histories, possibly due to AGN-driven feedback regulating or rejuvenating star formation. This difference indicates that the growth and evolutionary pathways of AGN hosts may not be as tightly linked to stellar mass as in non-AGN systems. The similar trend of the age with stellar mass can be found in the studies of \citet{mzr_Gallazzi2005} (see figure 8).


\subsubsection{Metallicity Vs Age}
While investigating the age vs metallicity of the galaxies, we found that for star-forming galaxies metallicity decrease with increasing age. This scenario is same for some of the Composites as well. But for most of the composites and AGNs, the metallicity increases with the age of the galaxies. In \citet{mzr_Gallazzi2005}, where galaxies are not separated into single or multi-nucleus systems, the age–metallicity relation is generally weak and driven mainly by stellar mass, younger galaxies tend to be more metal-poor, older galaxies somewhat more metal-rich, but with very large scatter, especially at low and intermediate masses. In contrast, our multi-nuclei sample shows a more structured behaviour. Star-forming galaxies in multi-nucleus systems clearly follow the classical trend seen in \citet{mzr_Gallazzi2005}, with metallicity decreasing as age increases. However, many Composite galaxies and almost all AGN-host galaxies in our sample show the opposite behaviour, exhibiting increasing metallicity with increasing age.

This suggests that the chemical enrichment histories of multi-nucleus systems, particularly those hosting AGN, may differ from the general galaxy population studied, likely due to merger driven gas inflows, bursty star formation, and AGN feedback, which can accelerate or regulate chemical evolution in ways not captured in mixed, non-categorized samples as used by \citet{mzr_Gallazzi2005}.



\subsection{Correlation between the physical properties of primary and secondary companions in dual systems}

\begin{figure}
    \centering
    \includegraphics[width=0.45\textwidth]{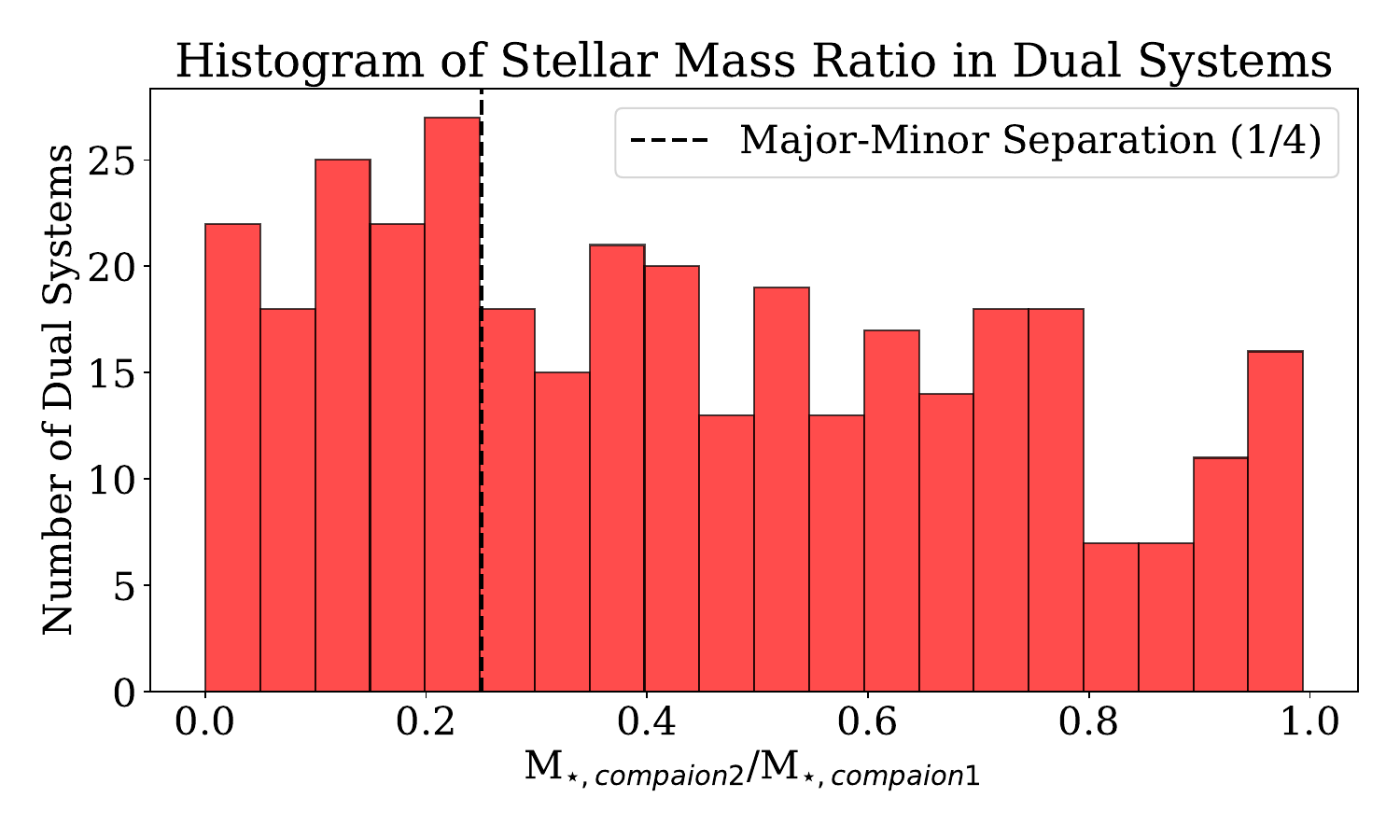}
    \caption{Distribution of ratio of stellar masses of Companion2 and Compaion1 of dual-nuclei systems}
    \label{fig:dual_mass_histo}
\end{figure}

\begin{figure*}
    \centering
    \begin{subfigure}[b]{0.5\textwidth}
         \centering
         \includegraphics[width=\textwidth]{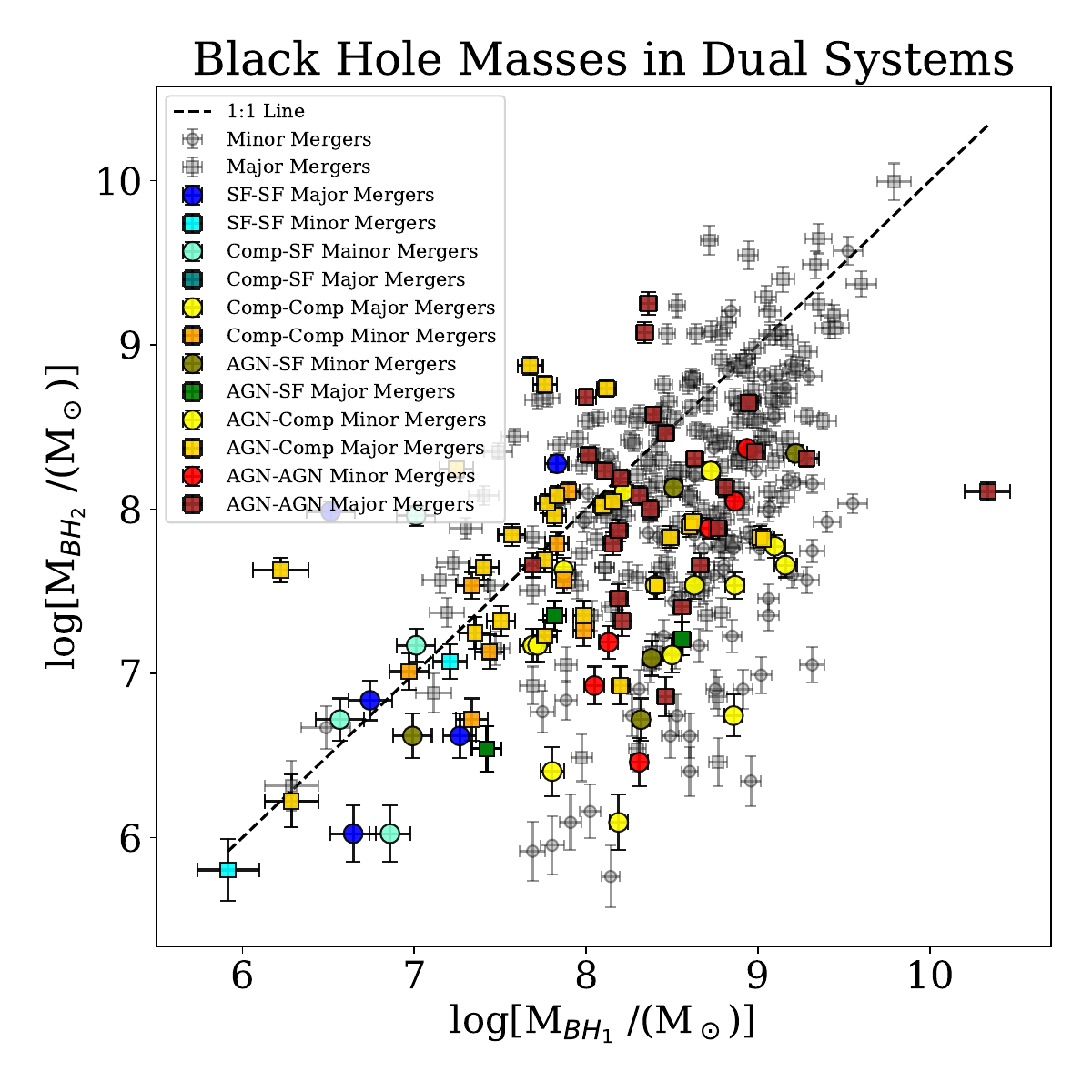}
     \end{subfigure}
     
     \hfil
     \begin{subfigure}[b]{0.24\textwidth}
         \centering
         \includegraphics[width=\textwidth]{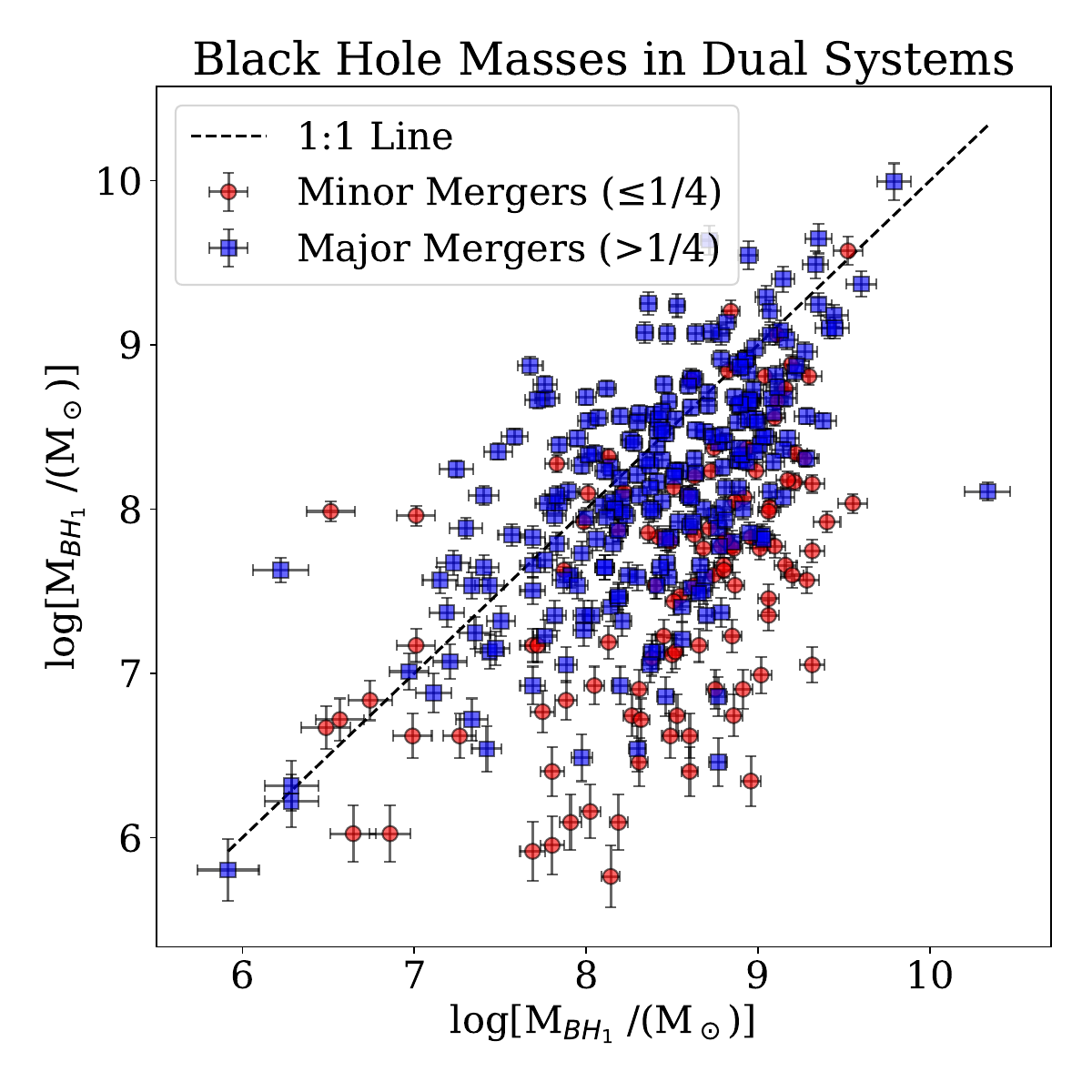}
     \end{subfigure}
     \hfil
     \begin{subfigure}[b]{0.24\textwidth}
         \centering
         \includegraphics[width=\textwidth]{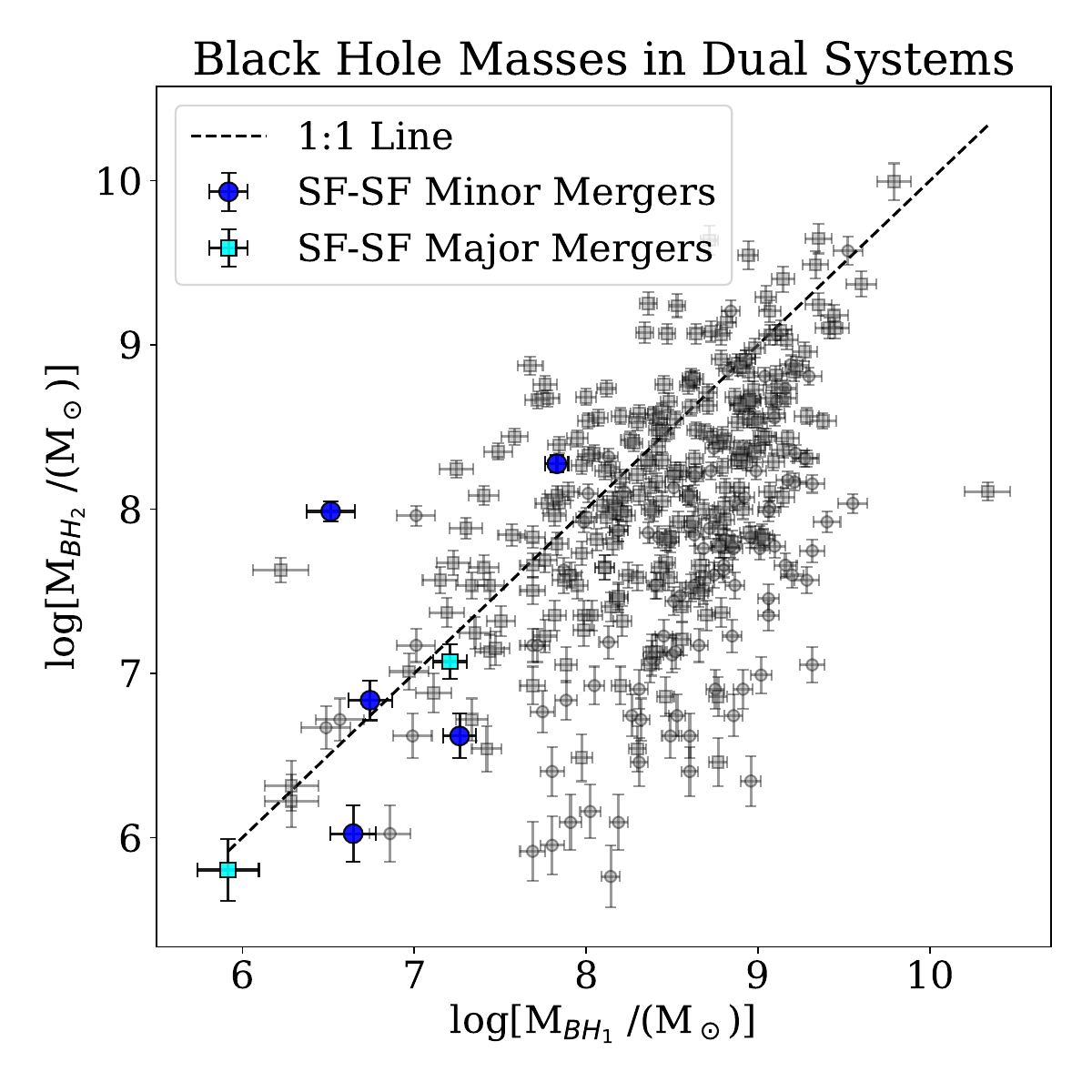}
     \end{subfigure}
     \hfil
     \begin{subfigure}[b]{0.24\textwidth}
         \centering
         \includegraphics[width=\textwidth]{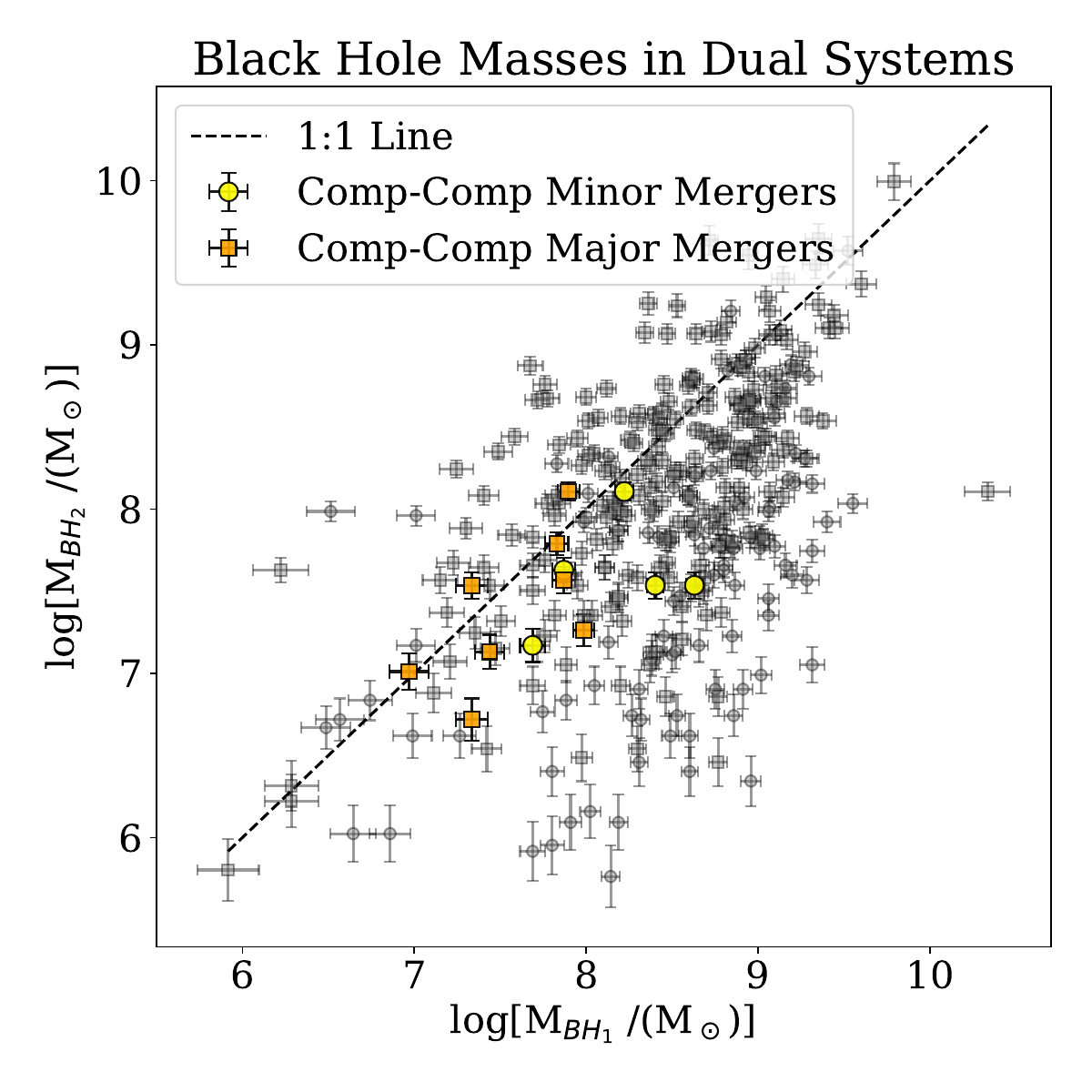}
     \end{subfigure}
     \hfil
     \begin{subfigure}[b]{0.24\textwidth}
         \centering
         \includegraphics[width=\textwidth]{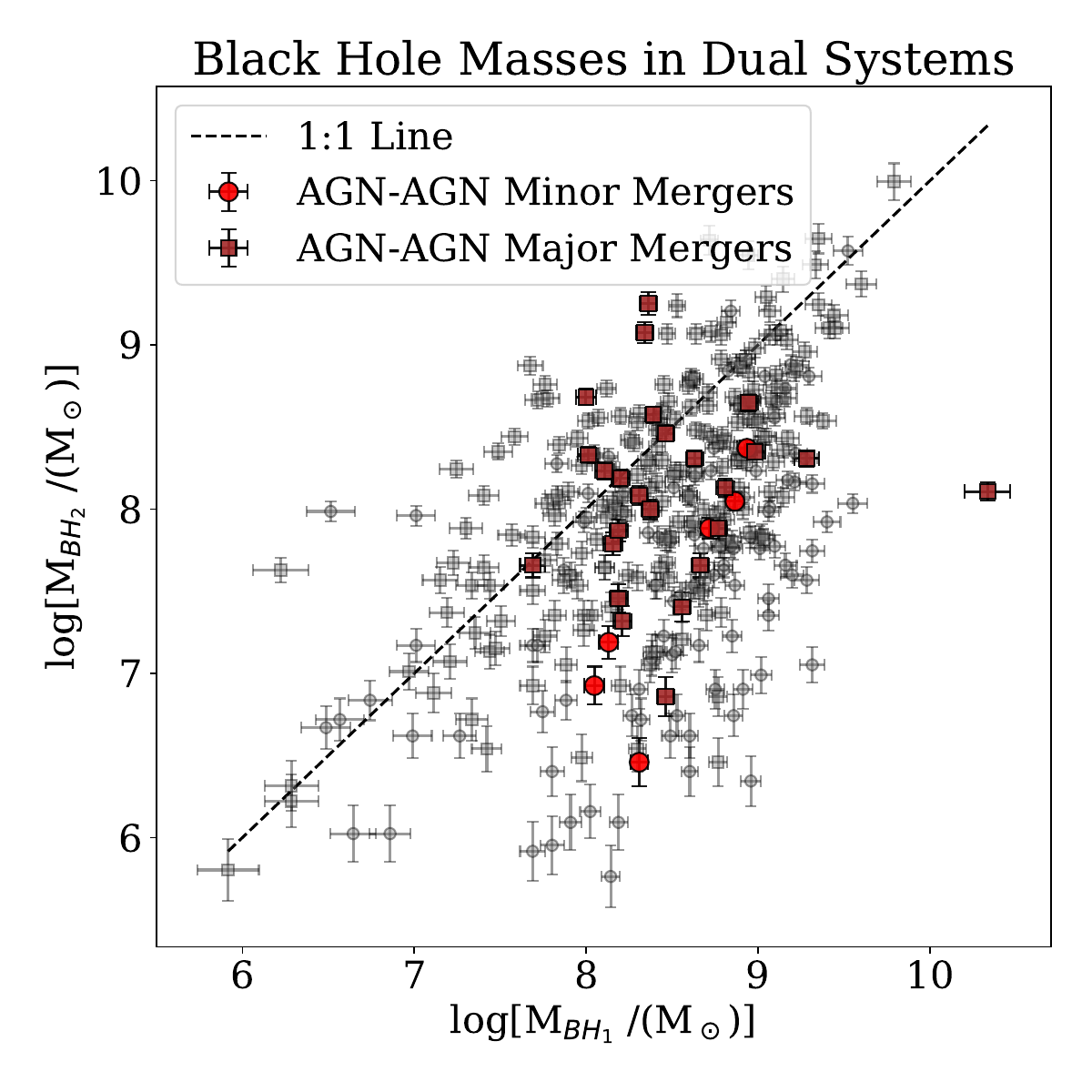}
     \end{subfigure}
     \hfil
     \begin{subfigure}[b]{0.24\textwidth}
         \centering
         \includegraphics[width=\textwidth]{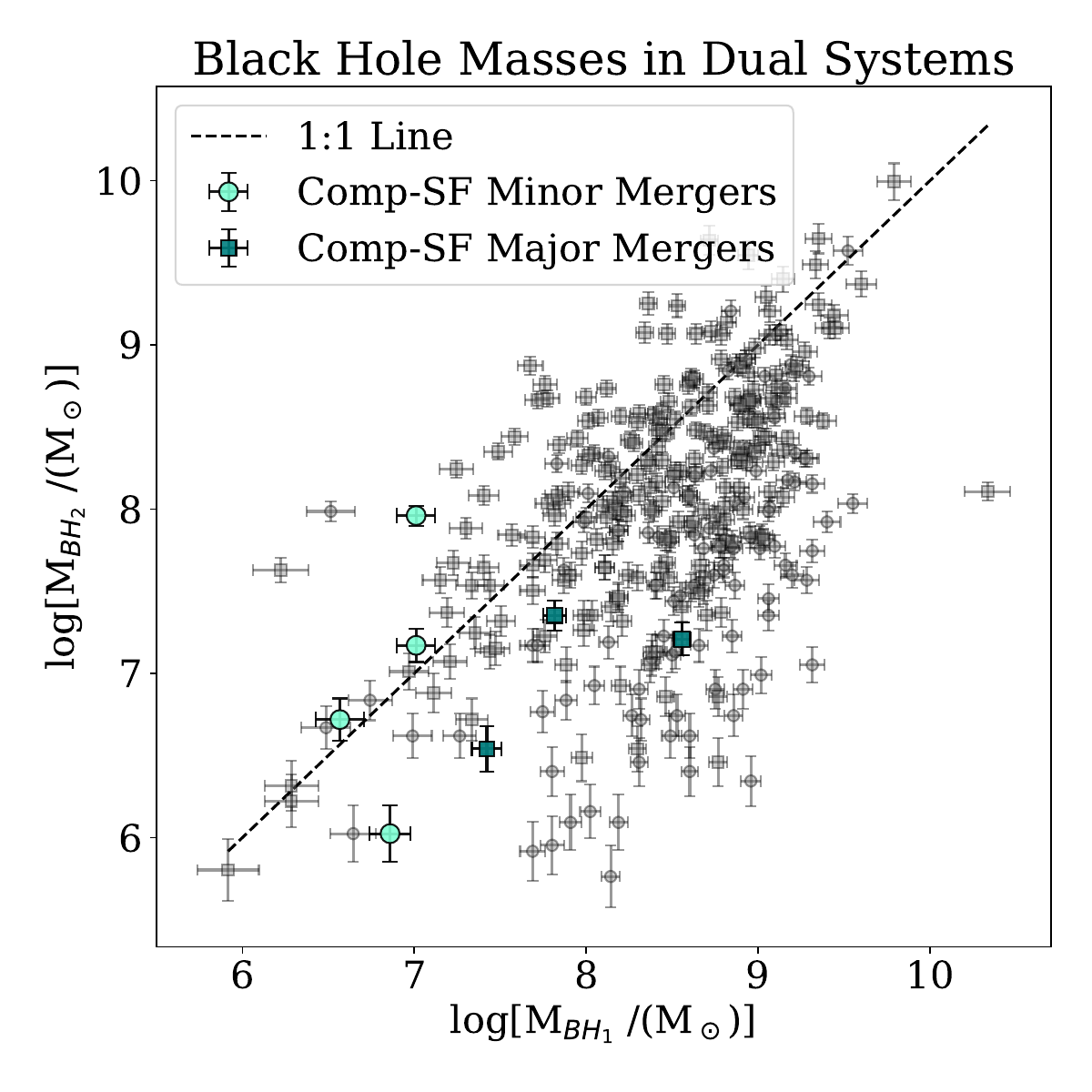}
     \end{subfigure}
     \hfil
     \begin{subfigure}[b]{0.24\textwidth}
         \centering
         \includegraphics[width=\textwidth]{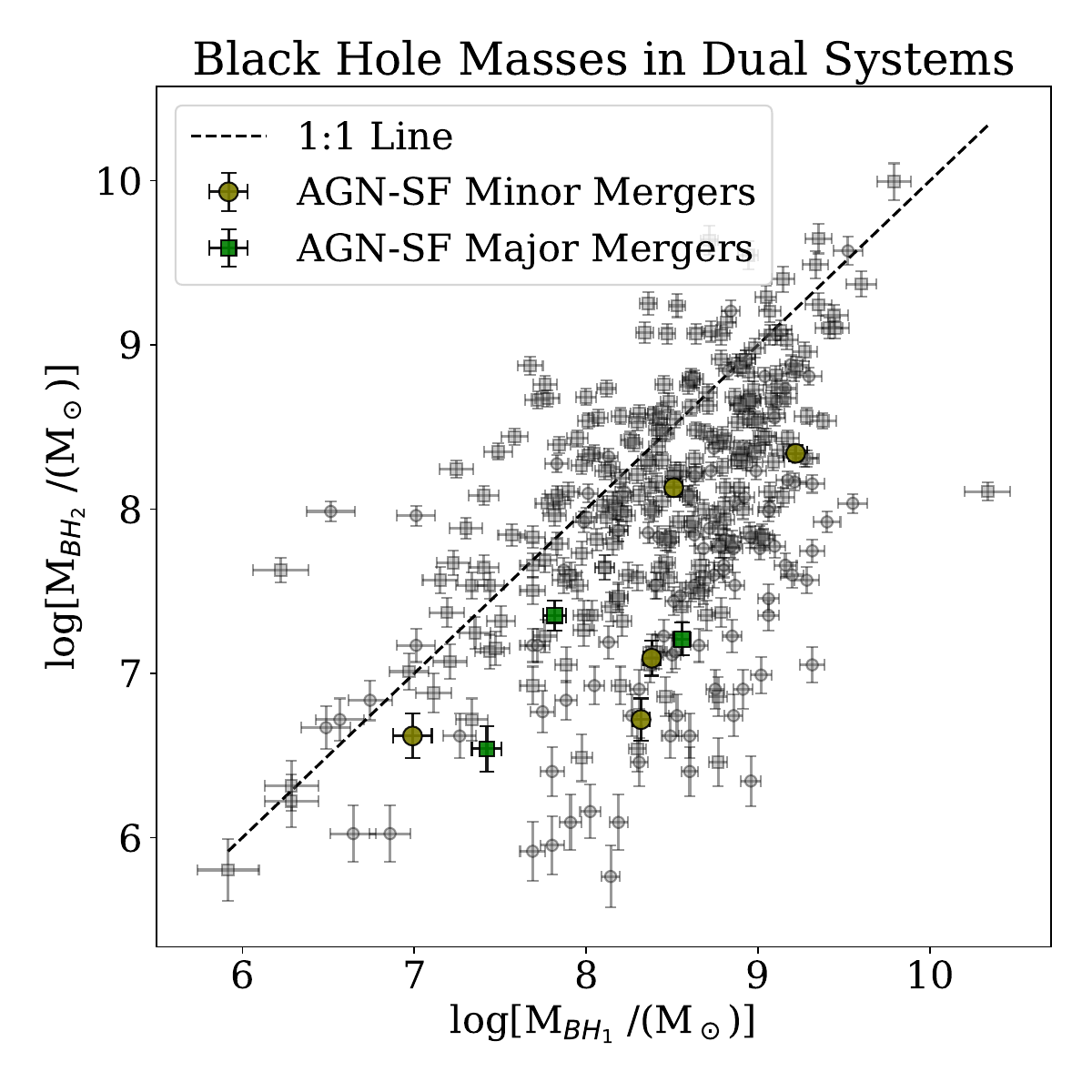}
     \end{subfigure}
     \hfil
     \begin{subfigure}[b]{0.24\textwidth}
         \centering
         \includegraphics[width=\textwidth]{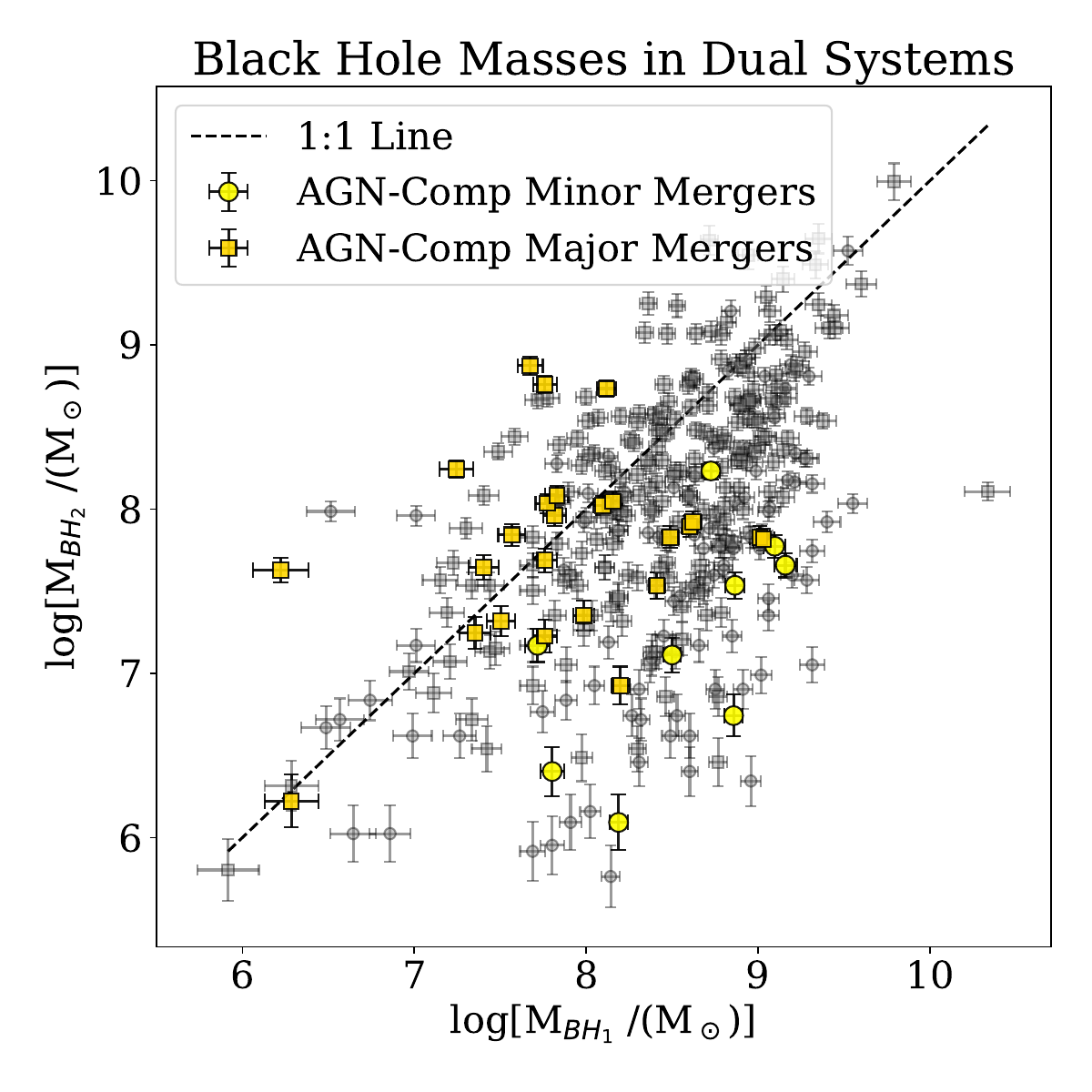}
     \end{subfigure}
    \caption{Black hole masses of Compaion2 Vs Companion1 for different combinations of systems for dual nuclei systems.}
    \label{fig:dual_mbh}
\end{figure*}

In this section, we investigate how paired nuclei co-evolve within the same host environment, whether their physical properties are correlated, and, if so, how these properties depend on each other. Hence, in this section, we explore the similarities and differences between the companions and discuss their implications on the dual-nucleus systems. 

As noted earlier, we identified 341 dual-nucleus systems in our sample for which both nuclei exhibit good-quality spectral fits, corresponding to a total of 682 individual sources. However, not all nuclei in these systems could be classified using the BPT diagram. A subset of 99 dual systems contains both nuclei successfully classified into one of the categories: star-forming (SF), Composite, AGN–LINER, or AGN–Seyfert. Within this classified subset of dual systems, we find a diverse range of nuclear pairings. A total of 31 systems consist of AGN–AGN pairs, here AGN refers collectively to AGN–LINER and AGN–Seyfert classifications. In addition, we identify 7 SF–SF systems in which both nuclei are purely star-forming, and 13 Composite–Composite systems where each nucleus shows mixed ionization properties. Mixed-type systems are also present: 8 AGN–SF systems, containing one AGN and one star-forming nucleus, along with 34 AGN–Composite systems, which form the most common mixed pairing in our sample. Finally, 6 Composite–SF systems are identified, comprising one Composite nucleus paired with a star-forming companion. Together, these classifications reveal the wide variety of nuclear combinations present in dual-nucleus galaxies and provide a basis for examining how the physical properties of paired nuclei relate to one another.

For each dual system, we labelled the two nuclei as \say{Companion 1} and \say{Companion 2} based on their stellar masses, designating the more massive nucleus as \say{Companion 1}. The distribution of the stellar mass ratio of these systems are shown in figure \ref{fig:dual_mass_histo}. We also classify these systems as undergoing major mergers (mass ratio $\geq 1/4$) or minor mergers (mass ratio $< 1/4$). We found that out of the 341 dual-nucleus systems, 106 dual systems ($\sim 31\%$) are undergoing Major mergers and 235 dual systems ($\sim 69\%$) are undergoing Minor mergers. Thus, the distribution indicates that the vast majority of dual systems correspond to minor mergers, with a smaller but noticeable extension toward more equal-mass, major-merger configurations. The distribution of major and minor mergers for each dual-nucleus pair type is summarised in Table~\ref{tab:major_minor_pairs_type}. The table shows that, across all categories, minor mergers dominate, consistent with the overall stellar-mass–ratio distribution of the full dual-nucleus sample.

\begin{table}
\centering
\begin{tabular}{lcc}
\hline
\textbf{Pair Type} & \textbf{Major Mergers} & \textbf{Minor Mergers} \\
\hline
AGN--AGN       & 6  & 25 \\
SF--SF         & 5  & 2  \\
Comp--Comp     & 5  & 8  \\
AGN--SF        & 5  & 3  \\
AGN--Comp      & 9  & 25 \\
Comp--SF       & 4  & 2  \\
\hline
\end{tabular}
\caption{Number of major and minor mergers for different dual-nucleus pair types.}
\label{tab:major_minor_pairs_type}
\end{table}

\subsubsection{Black hole masses of the dual nuclei systems}
Having established the merger classifications, we next investigate how the physical properties compare between the two nuclei in each system. As a first step, we examine the black hole mass ratio between the paired nuclei to understand how black hole growth proceeds within dual-nucleus galaxies. Figure~\ref{fig:dual_mbh} shows the distribution of the black hole mass of \say{Companion~1} versus that of \say{Companion~2}.

In the figure, we observe a general positive trend, i.e., the black hole mass of \say{Companion~2} tends to increase as the black hole mass of \say{Companion~1} increases. However, a significant number of systems deviate from the one-to-one line, exhibiting cases where $M_{\rm BH,1} < M_{\rm BH,2}$ even though $M_{\star,1} > M_{\star,2}$. A total of 169 dual systems fall into this regime, lying above the $M_{\rm BH,1} = M_{\rm BH,2}$ line. This behaviour is present across both major and minor mergers and appears in all pair categories, including AGN--AGN, SF--SF, Composite--Composite, AGN--SF, AGN--Composite, and Composite--SF systems. Thus, this trend is independent of merger type and does not depend on the nuclear classification of the galaxy pair.

\begin{figure}
    \centering
    \includegraphics[width=0.99\linewidth]{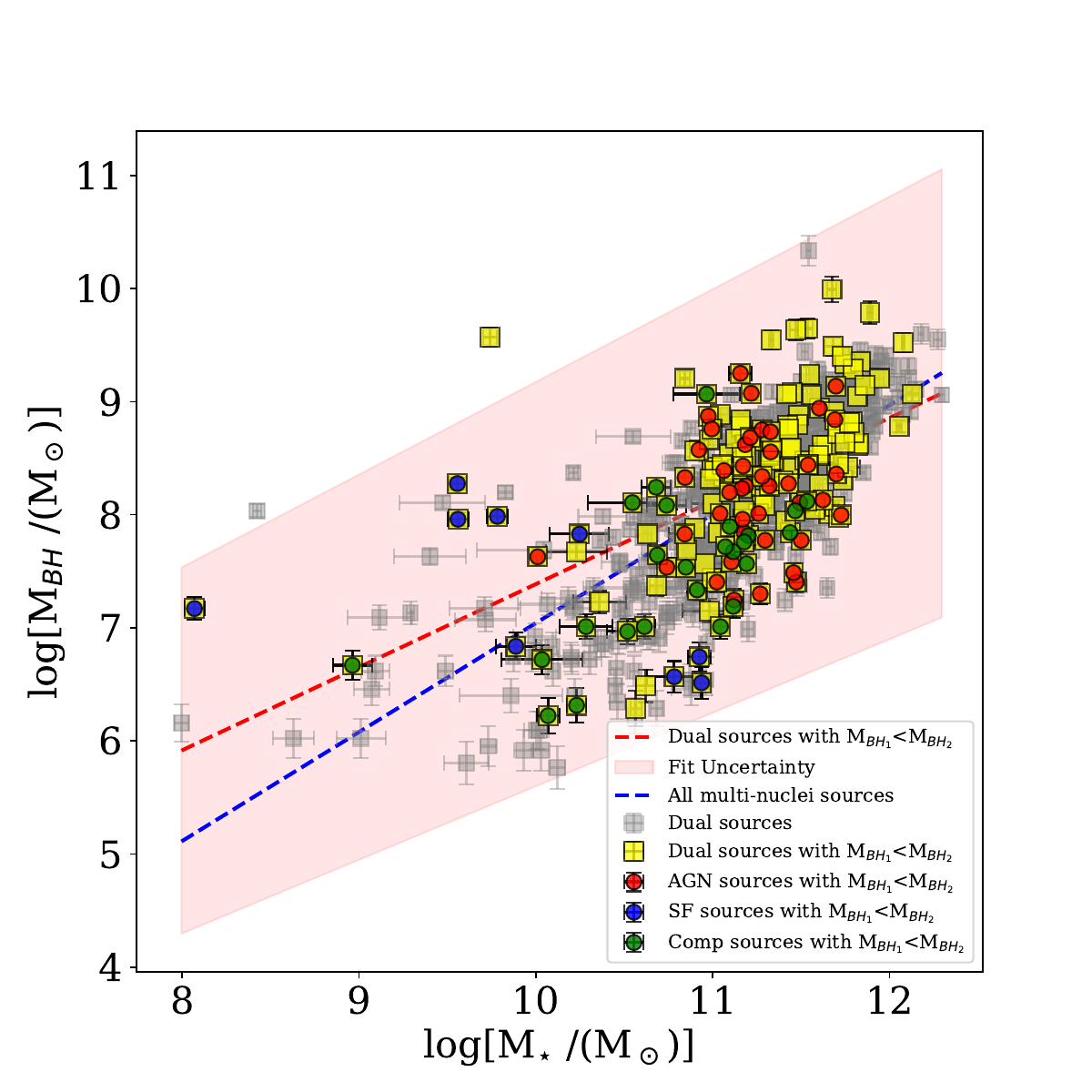}
    \caption{Black Hole mass Vs Stellar mass for 169 dual systems for which black hole mass of "Companion--2" is greater that "Companion--1".}
    \label{fig:mbh_mstar_dual21}
\end{figure}

To further assess whether these apparently \say{over-massive} black holes follow the expected global scaling relations, we examined the subset of systems for which $M_{\rm BH,1}/M_{\rm BH,2} < 1$ but $M_{\star,1}/M_{\star,2} > 1$. Fitting the $M_{\rm BH}$--$M_\star$ relation for this subset yields a slope and intercept of $a = 0.73 \pm 0.08$, and intercept $b = 0.04 \pm 0.94$, which is shallower than the relation obtained for the full multi-nuclei sample (subsection \ref{subsubsec:mbh_mstar}) with slope $0.96 \pm 0.03$ and intercept $-2.59 \pm 0.34$. Despite the difference in slope, the two relations remain statistically consistent within their uncertainties. As illustrated in Figure~\ref{fig:mbh_mstar_dual21}, the best-fitting relation for the full multi-nuclei sample lies within the $1\sigma$ confidence region around the fitted relation for the 169 dual systems. This demonstrates that, although these systems display an inverted black hole mass ordering relative to their stellar masses, their overall scaling with stellar mass remains compatible with the global $M_{\rm BH}$--$M_\star$ trend found for the entire multi-nucleus population.

\begin{figure*}
    \centering
    \begin{subfigure}[b]{0.33\textwidth}
    \centering
        \includegraphics[width=\textwidth]{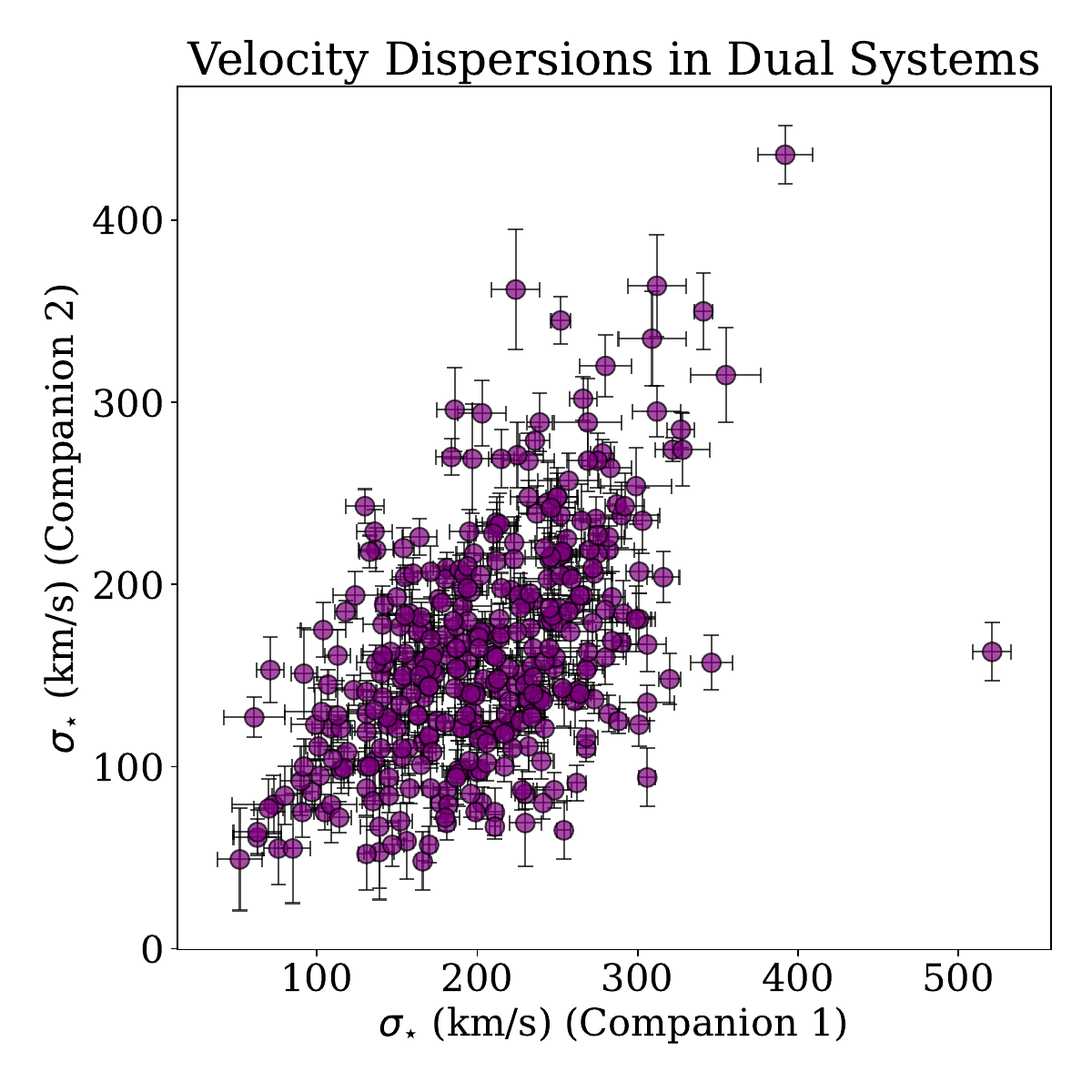}
        \caption{}
        \label{}
    \end{subfigure}
    \hspace{0.03\textwidth}
    \begin{subfigure}[b]{0.33\textwidth}
    \centering
        \includegraphics[width=\textwidth]{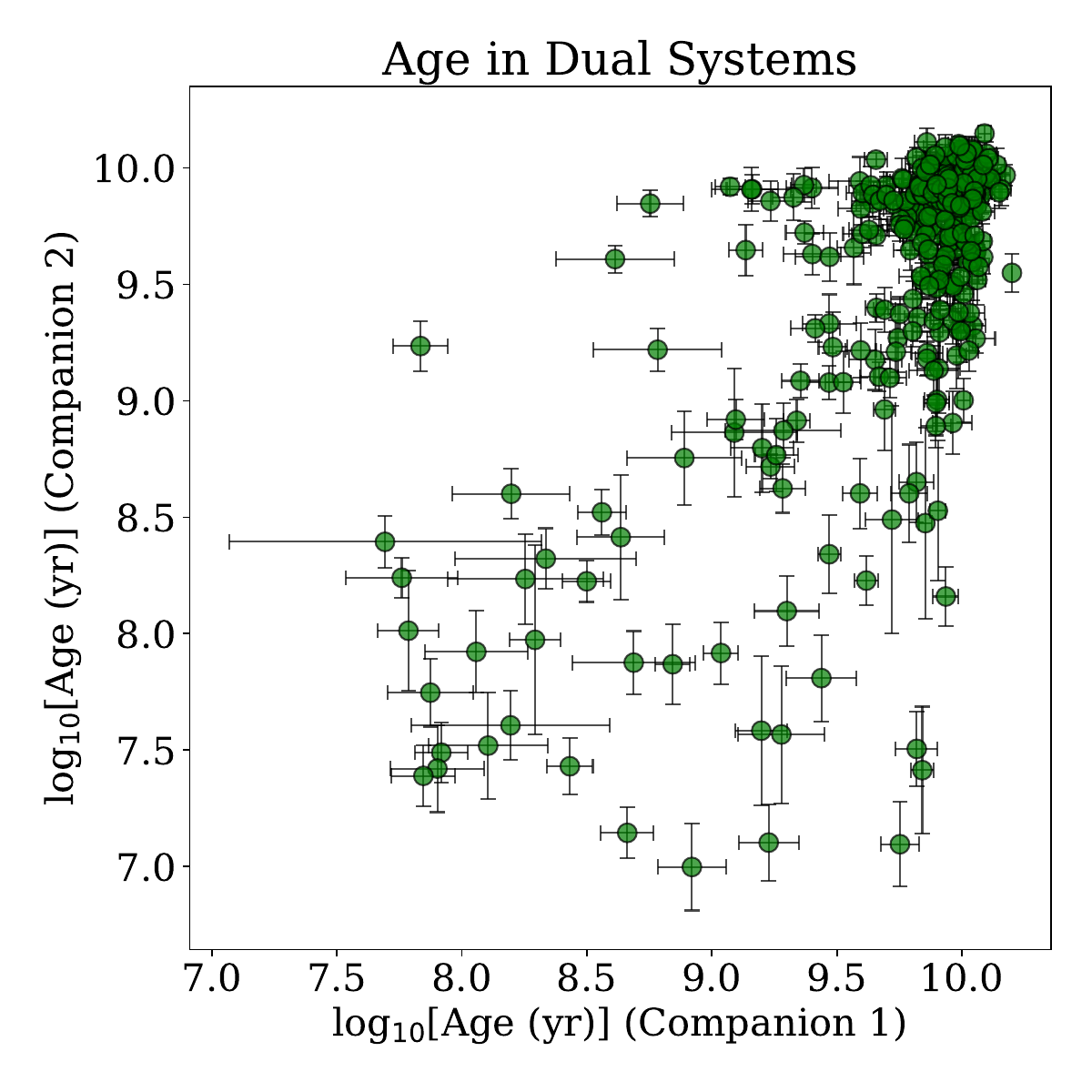}
        \caption{}
        \label{}
    \end{subfigure}   
    \vspace{0.1cm}
    \begin{subfigure}[b]{0.33\textwidth}
    \centering
        \includegraphics[width=\textwidth]{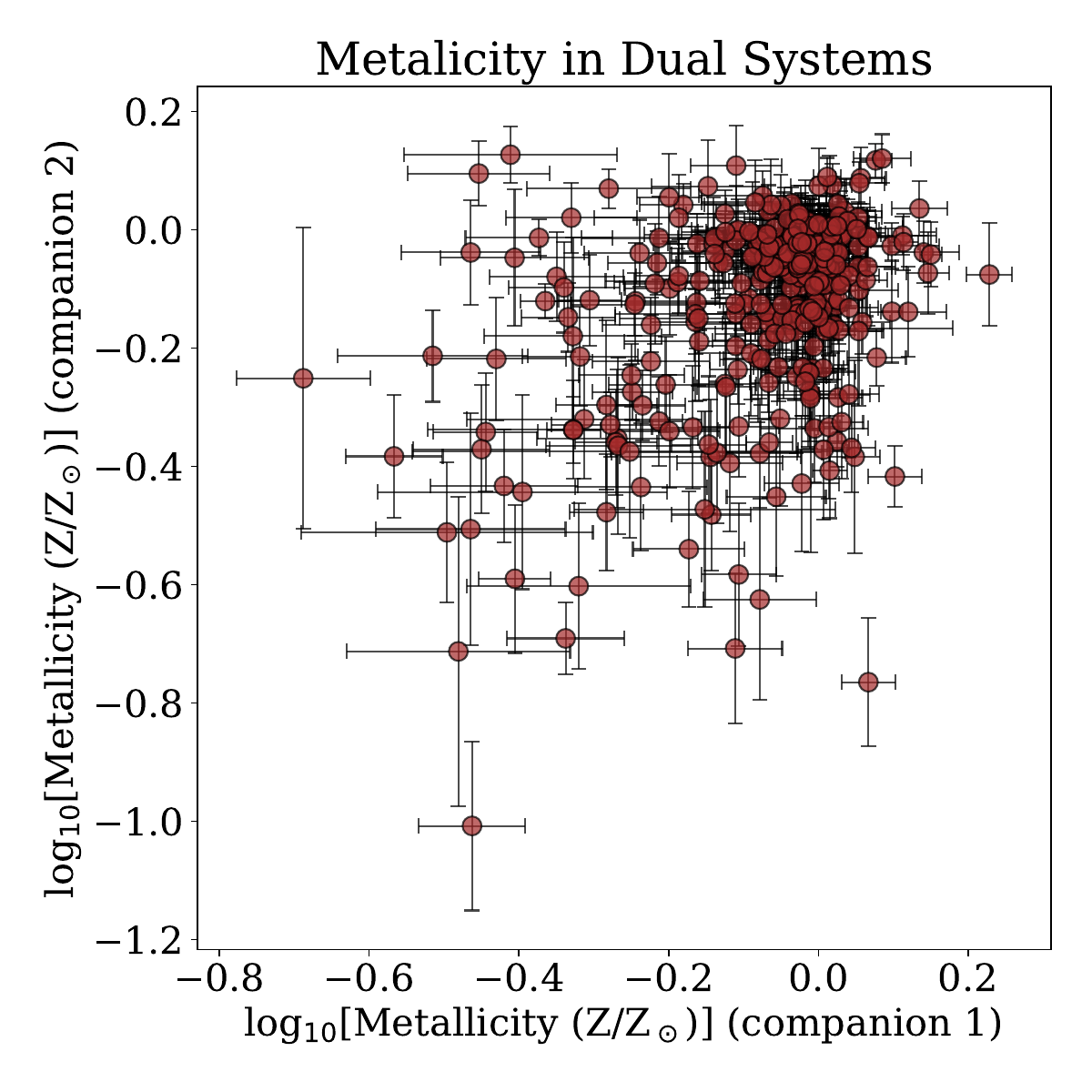}
        \caption{}
        \label{}
    \end{subfigure}
    \hspace{0.03\textwidth}
    \begin{subfigure}[b]{0.33\textwidth}
    \centering
        \includegraphics[width=\textwidth]{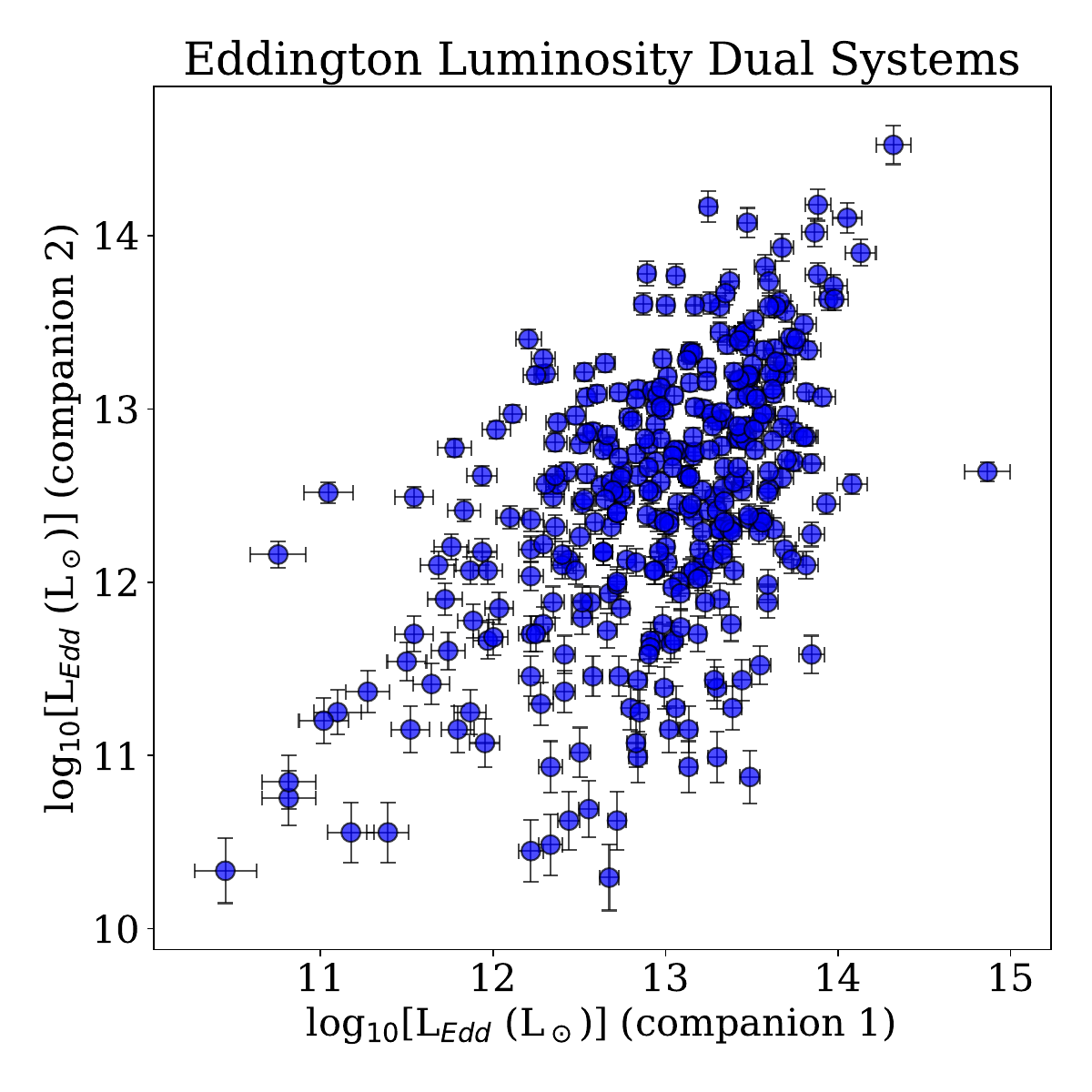}
        \caption{}
        \label{}
    \end{subfigure}
    \caption{Different properties of Companion1 Vs Companion2 of the dual nuclei systems}
    \label{fig:all_prop_comp1_vs_comp2}
\end{figure*}

\subsubsection{Other properties of the dual nuclei system}
As a next step, we explore how other physical properties, i.e., stellar dispersion velocity, age, metallicity and Eddington Luminosity of the two nuclei compare within each dual system (see Figure \ref{fig:all_prop_comp1_vs_comp2}). The velocity dispersion comparison shows a clear trend in which Companion~2 generally exhibits lower dispersion velocity than Companion~1, reflecting the expected dynamical differences between the two nuclei. In contrast, the age and metallicity comparisons display substantial scatter, with neither companion systematically older or more metal-rich. This indicates that the stellar population histories of the paired nuclei can differ significantly, even within the same interacting environment. Finally, the Eddington luminosity comparison reveals a more coherent pattern: the nucleus hosting the larger black hole typically exhibits the higher Eddington luminosity, consistent with the expected scaling between accretion luminosity and black hole mass. 

Overall, these results indicate that while dynamical properties such as velocity dispersion and accretion driven luminosity such as Eddington luminosity remain closely linked to the underlying mass hierarchy of the system, the stellar population properties of the two nuclei, i.e., age and metallicity are heterogeneous. This suggests that although both nuclei share the same global gravitational environment, their star-formation and chemical enrichment histories can diverge substantially. This may happen due to localized processes such as gas inflows, star formation, and AGN feedback.

\subsection{Black hole masses and the separation between two nuclei of the dual nuclei systems}
We additionally examined whether the Black hole mass of the dual nuclei systems shows any dependence on the projected separation between the two companions and whether this behaviour differs for major and minor mergers. Figure \ref{fig:mbh_separation} shows the ratio of masses of black holes of two companions in a dual system as a function of separation between them. From this figure we can see that the ratio of the mass of the black holes spans across all separation (~5–35 kpc) with substantial scatter. Both major and minor mergers systems are also distributed over all different measured projected separations and their black hole mass ratios.
No clear dependence of mass ratio on separation suggests that differential growth between the two black holes is not strongly driven by the stage of the merger, as traced by projected separation.

\begin{figure}
    \centering
    \includegraphics[width=0.45\textwidth]{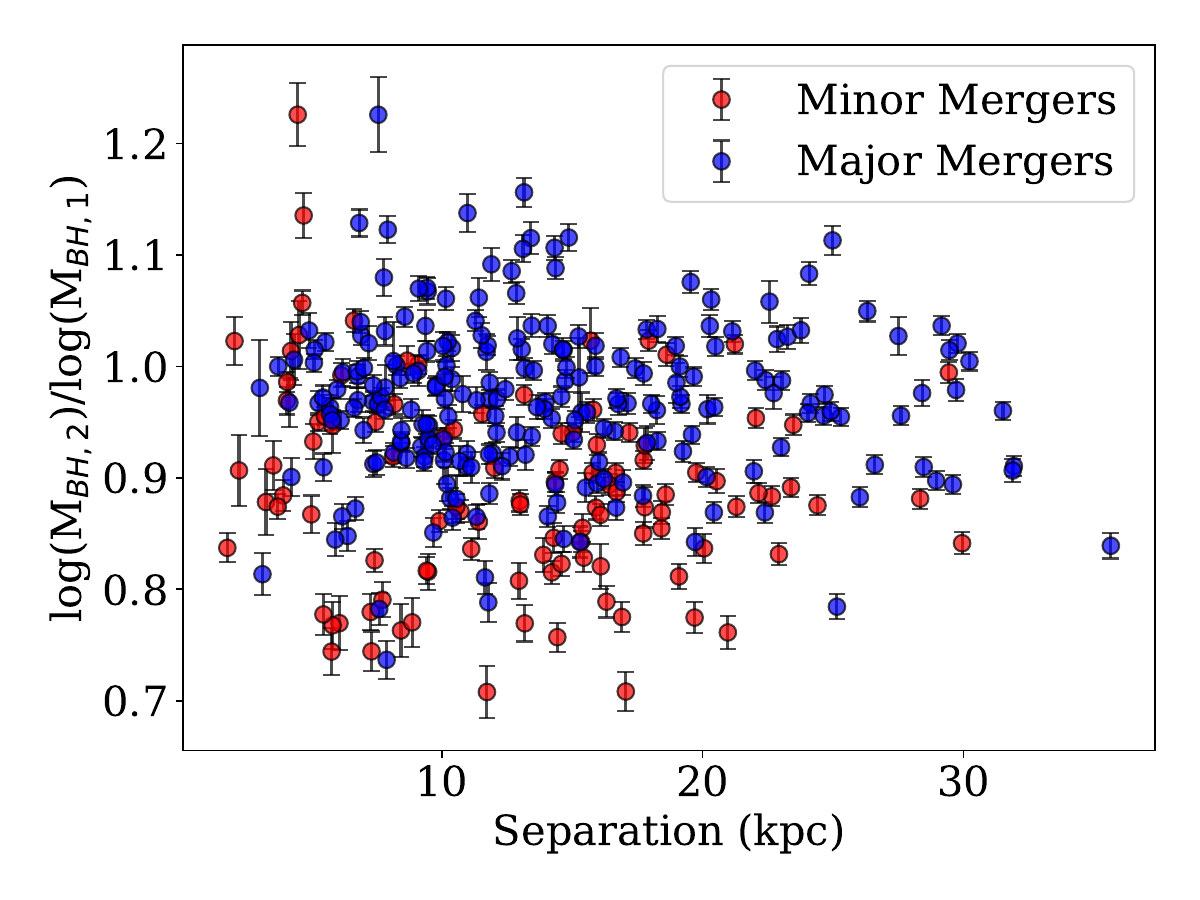}
    \caption{Ratio of black hole mass of Companion2 to Companion1 vs projected separation between them for the dual-nuclei system.  The red and blue circles represents the major and minor merger systems, respectively.}
    \label{fig:mbh_separation}
\end{figure}

\section{Discussion}
\label{sec:discussion}

This study presents a spectroscopic analysis of dual and multi-nucleus systems identified from the GOTHIC survey using SDSS data. By applying the \ppxf\ fitting technique to high-quality spectra, the work derives key physical properties such as stellar kinematics, stellar mass, age, metallicity, and supermassive black hole masses, etc. The analysis focuses primarily on dual-nucleus systems and compares their properties with those of single-nucleus systems to understand black hole growth, stellar population evolution, and the impact of galaxy mergers. 

It is to be noted that, for our sample, the redshift range is $z: [0.008$–$0.272]$, with a median of $z \sim 0.094$ and a peak around $z \sim 0.07$. The physical scale corresponding to the 3$^{\prime \prime}$ SDSS fibre varies significantly across this range: At the minimum redshift, the SDSS fibre corresponds to a physical scale of $\sim 0.5$ kpc,  while at the peak of the distribution of $z$ it is $\sim 4.0$ kpc. At the median redshift, the fibre covers $\sim 5.4$ kpc, increasing to $\sim 12.6$ kpc at the maximum redshift. This shows that, especially at higher redshifts, the SDSS fibre probes a substantial fraction of the galaxy, potentially including both nuclear and extended stellar components. 

Therefore, $\sigma_\star$ and other properties in our sample should be interpreted as aperture-averaged properties, rather than a purely central measurement. However, the spectra used in this work correspond to individual SDSS spectroscopic targets, i.e., distinct spectra for individual nuclei within multi-nucleus systems was taken, and the GOTHIC sample itself was visually confirmed to contain distinct nuclei. Therefore, the measured properties are expected to be dominated by the targeted nucleus. Nevertheless, for very close systems, particularly at higher redshifts, some contamination from the companion nucleus and its surrounding stellar light within the SDSS fibre cannot be completely excluded and may contribute to the observed scatter in the derived relations. 
Besides that, most of our sources lie near $z \sim 0.05$–0.15, where the fibre covers $\sim 3$–7.8 kpc. While this is larger than the strictly nuclear region, it is still dominated by the central galaxy potential, particularly for massive systems.  This aperture effect may introduce additional scatter in the measurement of different properties. While aperture effects may introduce some averaging at higher redshift, the relatively narrow redshift range and dominance of intermediate-redshift sources suggest that this does not significantly bias the main scaling relations.

In this study, we examined the correlations between nuclear and galaxy parameters of closely merging galaxies, including stellar mass, velocity dispersion, stellar age, metallicity, mass-to-light ratio, and black hole mass, and compared them to single-nuclei systems. We find significant differences which indicate that nuclear properties undergo changes during the merging process, and not just when the SMBHs coalesce.

The black hole mass–stellar mass relation in our sample broadly follows the well-known positive correlation, with AGN and Composite nuclei showing stronger trends than star-forming nuclei. AGN-LINERs exhibit a steeper slope, suggesting more efficient black hole growth in massive, evolved systems, while AGN-Seyferts show a shallower relation. However, the main result is that the SMBH masses of dual/multiple nuclei systems are significantly higher than that of single nuclei for a given stellar mass, clearly showing that SMBHs grow in mass during galaxy mergers. The growth is expected as a result of the infall of gas towards the centers of galaxies, leading to star formation and AGN activity. This has also been shown in simulations \citep{lin.etal.2023} and is expected from theories of hierarchical growth of SMBHs \citep{dimatteo.etal.2008,volonteri.etal.2020} There are also some outliers hosting unusually massive black holes relative to their stellar mass, which we will followup in subsequent studies.


Stellar population properties also reveal systematic differences. The mass–metallicity relation shows increasing metallicity with stellar mass, with AGN nuclei being the most metal-rich and star-forming nuclei the most metal-poor. Multi-nuclei systems show systematic offsets compared to classical single-nucleus relations, with higher metallicities at low masses and lower metallicities at high masses, likely due to merger-driven gas flows and interactions. Stellar age generally increases with mass, but the rate differs by spectral class: star-forming and Composite nuclei exhibit rapid age increases, whereas AGN hosts show a more gradual rise, indicating extended or regulated star-formation histories. The age–metallicity relation is also structured: star-forming nuclei follow the classical trend of decreasing metallicity with age, while most Composite and AGN nuclei show increasing metallicity with age, contrasting with the weak trends seen in single-nucleus samples. This suggests that chemical enrichment in multi-nucleus systems is strongly influenced by interactions, bursty star formation, and AGN feedback.


A key part of our analysis involved a classification of the dual nuclei systems. We find that they show a diverse mixture of AGN–AGN, Composite–Composite, star-forming pairs, and mixed-type pairs. Most systems are minor mergers, though a non-negligible number of major mergers are also present. Comparing the physical properties of the two nuclei provides insights into their co-evolution. We find that stellar velocity dispersion and Eddington luminosity remain ordered, Companion 1 (the more massive nucleus) generally exhibits higher values, as expected from their dependence on gravitational potential and black hole mass. However, stellar age and metallicity show no such ordering, with many systems displaying companions that differ substantially in their stellar population histories. This indicates that although both nuclei occupy the same merging environment, their star-formation and enrichment pathways can diverge significantly due to localized gas flows, bursty star formation, or AGN feedback.

The black hole masses of the two nuclei also show a positive correlation, but with notable exceptions: 169 systems exhibit $M_{\rm BH,2} > M_{\rm BH,1}$ despite $M_{\star,2} < M_{\star,1}$. These cases occur across AGN–AGN, Composite–Composite, mixed pairs, and both major and minor mergers, suggesting that asynchronous black hole growth is common in dual-nucleus galaxies. Nevertheless, these 169 systems follow an $M_{\rm BH}$–$M_\star$ relation consistent within uncertainties with that of the full sample, implying that deviations at the pair level do not break the global scaling relation.

Finally, we find no correlation between Eddington luminosity and separation between the nuclei, for either major or minor mergers. This indicates that the instantaneous accretion state of each black hole is not governed by the merger stage, but rather dominated by stochastic, small-scale fuelling processes.

\section{Conclusion}
\label{sec:conclusion}

Our study demonstrates that while multi-nucleus systems preserve many of the fundamental dynamical and black hole scaling relations observed in the broader galaxy population, they also show important differences compared to isolated single-nucleus galaxies. In particular, we find that supermassive black holes in dual/multi-nucleus systems are systematically more massive at a given stellar mass, suggesting enhanced black hole growth during the merger process even before the final coalescence of the black holes. Besides that, age and metallicity exhibit large pair-to-pair variations, although the significant scatter limits our ability to draw firm conclusions. However, it indicates the possibility of diverse evolutionary histories. We also identify several dual systems where the less massive stellar companion hosts the more massive black hole, indicating that black hole growth within interacting pairs may not always follow the stellar mass hierarchy, although these systems still remain within the intrinsic scatter of the global $M_{\rm BH}$--$M_\star$ relation. This highlights the importance of treating multi-nucleus systems as a distinct class when probing galaxy evolution, star-formation history, and SMBH growth in merging environments.

\section*{Acknowledgements}
We sincerely thank Dr. Michele Cappellari for his invaluable support, guidance, and insightful suggestions throughout this work. His input during multiple discussions greatly helped us improve the quality of the paper, particularly in running {\sc ppxf}, refining the fits, interpreting the results more robustly, and strengthening the scientific foundation of this study. Additionally, we thank Dr. Smitha Subramanian and Renu Devi for multiple helpful discussions related to this work.  Furthermore, MD and SB acknowledge the support of the Science and Engineering Research Board (SERB) Core Research Grant CRG/2022/004531, and the Department of Science and Technology (DST) grant DST/WIDUSHIA/PM/2023/25(G) for this research. This work has utilised the SDSS databases. For the Sloan Digital Sky Survey (SDSS), funding has been provided by the Alfred P. Sloan Foundation; the participating institutions are the National Aeronautics and Space Administration, the National Science Foundation, the U.S. Department of Energy, the Japanese Monbukagakusho, the Max Planck Society, and the Higher Education Funding Council for England.

\section*{Data Availability}

All the derived quantities and models produced in this study will be shared at reasonable request to the corresponding author.



\bibliographystyle{mnras}
\bibliography{example} 


\appendix
\section{Spectra with low quality fitting}
\label{sec:spec_bad}

As mentioned in subsection \ref{subsec:accptable_fits}, some of the fittings of the good-quality spectra were not good. They were categorised as quality flag = 3 and 4.  Figure \ref{fig:ppxf_fit_stellar_bad} shows the example of such sources.

\begin{figure}
     \centering
     \begin{subfigure}[b]{0.5\textwidth}
         \centering
         \includegraphics[width=\textwidth]{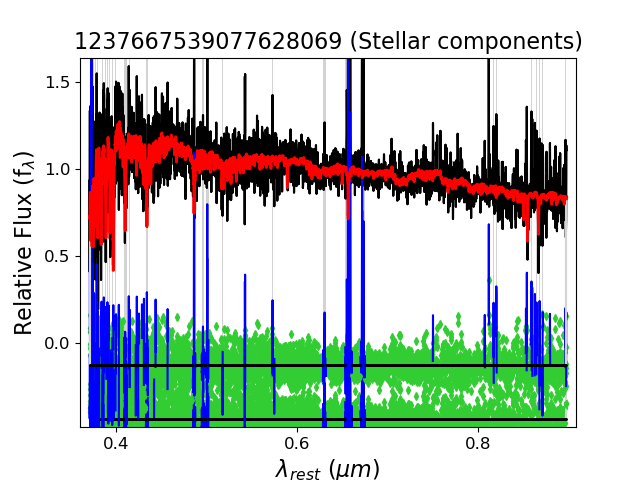}
         \caption{quality flag = 3}
         \label{fig:flag2}
     \end{subfigure}
     \hfill
     \begin{subfigure}[b]{0.5\textwidth}
         \centering
         \includegraphics[width=\textwidth]{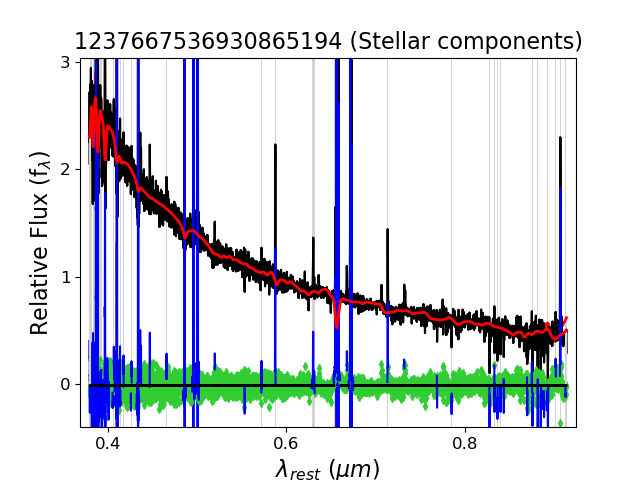}
         \caption{quality flag = 4}
         \label{fig:flag3}
     \end{subfigure}
        \caption{Comparison of the observed spectra with fitted model using stellar components only from {\sc pPXF} for two sources categorised as quality flag is 3 \& 4, respectively. The black and red points, respectively, represent the observed spectra and the fitted spectra. Green points denote the residuals from the fit, whereas blue points indicate those residuals that exceed the 3$\sigma$ threshold of the residual distribution. The grey vertical lines mark the pixels identified and clipped as outliers.}
        \label{fig:ppxf_fit_stellar_bad}
\end{figure}


\bsp	
\label{lastpage}


\end{document}